\documentclass[prd,amsmath,amssymb,aps,twocolumn]{revtex4-1}

\usepackage{color}

\newcommand{\subS}{\textrm{\tiny{S}}}
  \newcommand{\be}{\begin{equation} }
 \newcommand{\ee}{\end{equation}}
    \newcommand{\bes}{\begin{equation*} }
 \newcommand{\ees}{\end{equation*}}
  \newcommand{\bea}{\begin{eqnarray} }
 \newcommand{\eea}{\end{eqnarray}}
    \newcommand{\beas}{\begin{eqnarray*} }
 \newcommand{\eeas}{\end{eqnarray*}}
   \newcommand{\ba}{\begin{align} }
 \newcommand{\ea}{\end{align} }
  \newcommand{\bas}{\begin{align*} }
   \newcommand{\eas}{\end{align*} }

\usepackage{graphicx}
\usepackage{multirow}
\usepackage[colorlinks=true,citecolor=blue]{hyperref}

\begin{document}

\title{A Mode-Sum Prescription for Vacuum Polarization in Even Dimensions}

\author{Peter Taylor}
\email{peter.taylor@dcu.ie}
\affiliation{Center for Astrophysics and Relativity, School of Mathematical Sciences, Dublin City University, Glasnevin, Dublin 9, Ireland\\
 }
\author{Cormac Breen}
\email{cormac.breen@dit.ie}
\affiliation{School of Mathematical Sciences,\\ Dublin Institute of Technology, Kevin Street, Dublin 8, Ireland}

\date{\today}
\begin{abstract}
We present a mode-sum regularization prescription for computing the vacuum polarization of a scalar field in static spherically-symmetric black hole spacetimes in even dimensions. This is the first general and systematic approach to regularized vacuum polarization in higher even dimensions, building upon a previous scheme we developed for odd dimensions. Things are more complicated here since the even-dimensional propagator possesses logarithmic singularities which must be regularized. However, in spite of this complication, the regularization parameters can be computed in closed form in arbitrary even dimensions and for arbitrary metric function $f(r)$. As an explicit example of our method, we show plots for vacuum polarization of a massless scalar field in the Schwarzschild-Tangherlini spacetime for even $d=4,...,10$. However, the method presented applies straightforwardly to massive fields or to nonvacuum spacetimes.  \end{abstract}
\maketitle

\section{Introduction}
One particularly important approximation to a full theory of Quantum Gravity is semi-classical gravity, which is the treatment of quantum fields interacting with a classical spacetime metric via the semi-classical Einstein equations
\begin{align}
	G_{ab}=8\pi \langle T_{ab}\rangle.
\end{align}
Though there has been considerable debate on how exactly to interpret these equations, here we will have in mind the computation of the one-loop quantum correction about a particular fixed classical background. In this interpretation, the source-term on the right-hand side is the expectation value of the quantum stress-energy tensor operator for a particular field in a particular quantum state, where the stress-energy tensor operator is obtained by taking the classical expression for stress-energy for whatever fields are being considered and promoting the non-gravitational fields to operators. This procedure can also be adapted to gravitational perturbations about a fixed classical metric. This quantization procedure immediately leads to problems, namely that the source term after quantization is quadratic in an operator-valued distribution and hence divergent. A formal prescription to regularize the stress-energy tensor, the point-splitting scheme, dates back to DeWitt and Christensen \cite{dewitt1975quantum, ChristensenPointSplit}. Effectively, the prescription amounts to considering the stress-tensor evaluated at two nearby spacetime points and then subtracting a parametrix that encodes all the geometrical divergences in the coincident limit. This problem is already there in Minkowski spacetime where the familiar normal-ordering cure is equivalent to the point-splitting scheme applied to flat spacetime. In curved spacetime, however, there is no preferred vacuum state and hence a normal ordering procedure cannot be adopted.

The Christensen-DeWitt point-splitting scheme offers a formal resolution to the problem of regularization, but it does not inform how to compute regularized quantities in practice. Applying the point-splitting scheme in a way that is amenable to numerical evaluation has proven extremely difficult. The first significant breakthrough in this direction was the seminal work of Candelas and Howard \cite{CandelasHowardPhi2} who computed regularized vacuum polarization for a scalar field in the Schwarzschild black hole spacetime. Despite serious drawbacks, including its crucial dependence on WKB methods (which are problematic in the Lorentzian sector and generally ill-behaved near horizons) and lack of numerical efficiency, the Candelas-Howard approach has remained more or less the standard prescription for several decades.

In Ref.~\cite{TaylorBreenOdd}, we derived a new regularization scheme that was both conceptually clearer and much more efficient than the Candelas Howard approach. There are several other advantages, among them is the fact that the method is mostly agnostic to number of dimensions. This is quite remarkable given that the severity of the singularities to be regularized increases with the number of dimensions. In that paper, we treat only  the vacuum polarization of scalar fields in odd dimensional static spherically-symmetric spacetimes. In this paper, we extend the method to even dimensions also. Things are more complicated in even dimensions since the Feynman Green function possesses logarithmic singularities. We illustrate the power of the method by showing results for vacuum polarization for a scalar field in Schwarzschild-Tangherlini spacetimes with $d=4,6,8,10$. So efficient is the evaluation that the plots shown are generated on a standard laptop computer and are accurate to approximately 8-10 decimal places with only a few tens of modes. We note also that there is no conceptual obstacle to extend the methods presented in this series of papers to the computation of the regularized stress-energy tensor, though the calculations are much more involved.

Finally, apart from presenting this new method for even dimensions, the results presented for Schwarzschild-Tangherlini spacetimes are the first exact numerical results for vacuum polarization on the exterior of higher even-dimensional black hole spacetimes in the literature, at least as far as the authors are aware. There has been some work on regularized vacuum polarization and regularized stress-energy tensors in higher dimensions in the literature, but the focus has not been on computing exact quantities in a particular spacetime of interest but rather on general properties, expansions of the singular two-point function or approximations (see Refs.~\cite{Morgan:2007hp, DecaniniFolacciHadamardRen, Thompson:2008bk, MatyjasekHD1, MatyjasekHD2} for example).

\section{The Euclidean Green Function}
We consider a quantum scalar field on a static, spherically symmetric black hole spacetime of the form
\begin{align}
	\label{eq:metric}
	ds^{2}=-f(r)dt^{2}+dr^{2}/f(r)+r^{2}d\Omega^{2}_{d-2},
\end{align}
where $d\Omega^{2}_{d-2}$ is the metric on $\mathbb{S}^{d-2}$. Assuming the field is in a Hartle-Hawking state, we can adopt Euclidean techniques to simplify the problem. In particular, performing a Wick rotation $t\to-i\,\tau$ results in the Euclidean metric
\begin{align}
	ds^{2}=f(r)d\tau^{2}+dr^{2}/f(r)+r^{2}d\Omega_{d-2}^{2}.
\end{align}
It can be shown that this metric would possess a conical singularity unless we enforce the periodicity $\tau=\tau+2\pi/\kappa$ where $\kappa$ is the surface gravity. This discretizes the frequency spectrum of the field modes which now satisfy an elliptic wave equation
\begin{align}
 (\Box-m^{2}-\xi\,R)\phi=0,
\end{align}
where here and throughout $\Box$ is the d'Alembertian operator with respect to the Euclidean metric, $m$ is the scalar field mass and $\xi$ is the constant that couples the scalar to the gravitational field. The corresponding Euclidean Green function has the following mode-sum representation
\begin{align}
	\label{eq:modeGreen}
	G(x,x')=\frac{\kappa}{2\pi}\sum_{n=-\infty}^{\infty}e^{i n \kappa \Delta\tau}\sum_{l=0}^{\infty}\frac{(l+\mu)}{\mu\,\Omega_{d-2}}C_{l}^{\mu}(\cos\gamma)g_{nl}(r,r')
\end{align}
where $\mu=(d-3)/2$ and $\Omega_{d-2}=2\,\pi^{\mu+1}/\Gamma(\mu+1)$, $C_{l}^{\mu}(x)$ is the Gegenbauer polynomial and $\gamma$ is the geodesic distance on the $(d-2)$-sphere. The radial Green function satisfies
\begin{align}
	\label{eq:radialeqn}
	\Bigg[\frac{d}{dr}\Big(r^{d-2}f(r)\frac{d}{dr}\Big)-r^{d-2}\Big(\frac{n^{2}\kappa^{2}}{f(r)}+m^{2}+\xi\,R(r)\Big)\nonumber\\
	-r^{d-4}l(l+d-3)\Bigg]g_{nl}(r,r')=-\delta(r-r').
\end{align}
The solution can be expressed as a normalized product of homogeneous solutions
\begin{align}
	\label{eq:radialgreenfn}
	g_{nl}(r,r')=N_{nl}\,p_{nl}(r_{<})q_{nl}(r_{>}),
\end{align}
where $p_{nl}(r)$ and $q_{nl}(r)$ are homogeneous solutions which are regular on the horizon and the outer boundary (usually spatial infinity), respectively. We have adopted the notation $r_{<}\equiv \min\{r,r'\}$, $r_{>}\equiv \max\{r,r'\}$. The normalization constant is given by
\begin{align}
	\label{eq:norm}
	N_{nl}\,W\{p_{nl}(r),q_{nl}(r)\}=-\frac{1}{r^{d-2}f(r)},
\end{align}
where $W\{p,q\}$ denotes the Wronskian of the two solutions.

Now, computing the vacuum polarization involves taking the so-called coincidence limit $x'\to x$ of the Euclidean Green function. However, the mode-sum expression (\ref{eq:modeGreen}) does not converge in this limit reflecting the fact that the Green function satisfies a wave equation with a delta distribution source. In order to make the coincidence limit meaningful, we must regularize the mode-sum in a way described in the remainder of this paper.

\section{The Singular Propagator}
In the point-splitting approach to computing the regularized vacuum polarizations in the Hartle-Hawking state, one subtracts an appropriate two-point function from the Euclidean Green function and then one takes the coincidence limit $x'\to x$. Of course, in order for this to be meaningful, the short-distance singularities in the Green function and in the two-point function that one subtracts must cancel. This is guaranteed if the two-point function is chosen to be a parametrix for the scalar wave equation, that is, if we subtract a two-point function $G_{\subS}(x,x')$ that satisfies
\begin{align}
	(\Box-m^{2}-\xi\,R)G_{\subS}(x,x')=\delta(x,x')+S(x,x')
\end{align}
where $\delta(x,x')$ is the $d$-dimensional covariant delta distribution and $S(x,x')$ is an arbitrary smooth biscalar. There are other constraints on $G_{\subS}(x,x')$, the most important of which is that it must depend only on the geometry via the metric and its derivatives. This guarantees that the divergences in the semi-classical equations are renormalizable \cite{WaldQFT}. These criteria still only fix $G_{\subS}(x,x')$ up to the addition of an arbitrary smooth biscalar that depends functionally only on the geometry. So a choice must be made. Here, we adopt the Hadamard regularization prescription (see, e.g., \cite{DecaniniFolacciHadamardRen}), i.e., we define our singular propagator to be a Hadamard parametrix (or rather a one-parameter family of parametrices). In even dimensions, this is given by
\begin{align}
	G_{\subS}(x,x')=\frac{\Gamma(\frac{d}{2}-1)}{2 (2\pi)^{d/2}}\Bigg\{\frac{U(x,x')}{\sigma(x,x')^{\frac{d}{2}-1}}+V(x,x')\log(2\sigma/\ell^{2})\Bigg\}.
\end{align}
The biscalar $\sigma(x,x')$ is the world function with respect to the Euclideanized metric. The parameter $\ell$ is a lengthscale required to make the argument of the $\log$ dimensionless. The biscalars $U(x,x')$ and $V(x,x')$ are smooth and symmetric in their arguments. For a scalar field, $U(x,x')$ satisfies the wave equation
\begin{align}
	\sigma(\Box-m^{2}-\xi\,R)U&=(d-2)\sigma^{a}\nabla_{a}U\nonumber\\
	&-(d-2)U\,\Delta^{-1/2}\sigma^{a}\nabla_{a}\Delta^{1/2},
\end{align}
where $\sigma^{a}\equiv \nabla^{a}\sigma$ and $\Delta(x,x')$ is the Van Vleck Morrette determinant. Assuming the Hadamard ansatz for a series solution
\begin{align}
	U(x,x')=\sum_{p=0}^{\frac{d}{2}-2}U_{p}(x,x')\,\sigma^{p},
\end{align}
it can be shown that each coefficient $U_{p}(x,x')$ satisfies
\begin{align}
	\label{eq:HadamardUp}
	&(p+1)(2p+4-d)U_{p+1}+(2p+4-d)\sigma^{a}\nabla_{a}U_{p+1}\nonumber\\
	&-(2p+4-d)U_{p}\Delta^{-1/2}\sigma^{a}\nabla_{a}\Delta^{1/2}\nonumber\\
	&+(\Box-m^{2}-\xi\,R)U_{p}=0,
\end{align}
with boundary condition $U_{0}=\Delta^{1/2}$. The biscalar $V(x,x')$ satisfies the homogenous wave equation
\begin{align}
	\label{eq:Veqn}
	(\Box-m^{2}-\xi\,R)V(x,x')=0,
\end{align}
and since it is symmetric, it is also a solution with respect to the wave operator at $x'$. The Hadamard ansatz for $V(x,x')$,
\begin{align}
	V(x,x')=\sum_{p=0}^{\infty}V_{p}(x,x')\,\sigma^{p}
\end{align}
can be substituted into (\ref{eq:Veqn}) to obtain a sequence of recursion relations for $V_{p}$:
\begin{align}
	\label{eq:HadamardVp}
	(p+1)(2p+d)V_{p+1}+2(p+1)\sigma^{a}\nabla_{a}V_{p+1}\nonumber\\
	-2(p+1)V_{p+1}\Delta^{-\frac{1}{2}}\sigma^{a}\nabla_{a}\Delta^{\frac{1}{2}}\nonumber\\
	+(\Box-m^{2}-\xi\,R)V_{p}=0,
\end{align}
along with the boundary condition
\begin{align}
	(d-2)V_{0}+2\sigma^{a}\nabla_{a}V_{0}-2 V_{0}\Delta^{-\frac{1}{2}}\sigma^{a}\nabla_{a}\Delta^{\frac{1}{2}}\nonumber\\
	+(\Box-m^{2}-\xi\,R)U_{\frac{d}{2}-2}=0.
\end{align}

Subtracting the Hadamard parametrix formally regularizes the Euclidean Green function. However, most of the difficulty in the calculation is how to subtract this in a way that the coincidence limit can be taken in a meaningful way and such that the resultant expression is numerically tractable. The problem is that the divergences in the mode-sum representation of the Euclidean Green function manifest as the non-convergence of that mode-sum at coincidence, while the divergence in the Hadamard parametrix is explicitly geometrical. One must attempt to express the Hadamard parametrix as a mode-sum of the same form as the Euclidean Green function and then subtract mode-by-mode. This is the crux of the regularization problem in QFT in curved spacetime.

In order to attain the desired mode-sum representation of the Hadamard parametrix, we must expand in a judiciously chosen set of variables. Now, the world function possesses a standard coordinate expansion which to lowest order is simply $\sigma=\tfrac{1}{2}g_{ab}\Delta x^{a}\Delta x^{b}+\mathcal{O}(\Delta x^{3})$. Following Ref.~\cite{TaylorBreenOdd}, we eschew this standard expansion and instead assume an expansion of the form
\begin{align}
	\label{eq:SigmaExp}
	\sigma=\sum_{ijk}\sigma_{ijk}(r)w^{i}\Delta r^{j}s^{k}
\end{align}
where 
\begin{align}
	\label{eq:ExpVar}
	w^{2}=\frac{2}{\kappa^{2}}(1-\cos \kappa\Delta\tau),\qquad s^{2}=f(r)\,w^{2}+2 r^{2}(1-\cos\gamma).
\end{align}
We will refer to $w$ and $s$ as ``extended coordinates'' and we will formally treat these as $\mathcal{O}(\epsilon)\sim\mathcal{O}(\Delta x)$ quantities. Substituting this into the defining equation $\sigma_{a}\sigma^{a}=2\sigma$ and equating order by order uniquely determines the coefficients $\sigma_{ijk}(r)$. To leading order, we simply have $\sigma=\tfrac{1}{2}\epsilon^{2}(s^{2}+\Delta r^{2}/f)+\mathcal{O}(\epsilon^{3})$, where we insert explicit powers of $\epsilon$ as a book-keeping mechanism for tracking the order of each term. Analogous expansions may be assumed for $U_{p}(x, x')$ and $V_{p}(x,x')$,
\begin{align}
	\label{eq:UVExp}
	U_{p}(x,x')&=\sum_{ijk}u^{(p)}_{ijk}(r)\epsilon^{i+j+k}w^{i}\Delta r^{j}s^{k}\nonumber\\
	V_{p}(x,x')&=\sum_{ijk}v^{(p)}_{ijk}(r)\epsilon^{i+j+k}w^{i}\Delta r^{j}s^{k},
\end{align}
and substituting these into (\ref{eq:HadamardUp}) and (\ref{eq:HadamardVp}) determines the coefficients $u^{(p)}_{ijk}(r)$ and $v^{(p)}_{ijk}(r)$, respectively.

Combining (\ref{eq:SigmaExp}) and (\ref{eq:UVExp}) gives a series expansion for the Hadamard parametrix in terms of the expansion parameters $w$, $s$ and $\Delta r$. This type of computation is ideally suited to a symbolic computer package such as Mathematica. Since we are ultimately interested in the coincidence limit, let us simplify by taking the partial coincidence limit $\Delta r=0$, then it can be shown that %$U/\sigma^{d/2-1}+V\,\log(2\,\sigma/\ell^{2})$ 
the Hadamard parametrix possesses an expansion of the form
\begin{widetext}
\begin{align}
	\label{eq:HadamardExp}
	\frac{U}{\sigma^{\frac{d}{2}-1}}&+V\,\log(2\sigma/\ell^{2})=\sum_{i=0}^{ \frac{d}{2}+m-2}\sum_{j=0}^{ i}\mathcal{D}^{(+)}_{ij}(r)\epsilon^{2i-2\mu-1}\frac{w^{2i+2j}}{s^{2\mu+2j+1}}+\sum_{i=1}^{ \frac{d}{2}+m-2}\sum_{j=1}^{ i}\mathcal{D}^{(-)}_{ij}(r)\epsilon^{2i-2\mu-1}\frac{w^{2i-2j}}{s^{2\mu-2j+1}}\nonumber\\
	&+\log(\epsilon^{2}\,s^{2}/\ell^{2})\sum_{i=0}^{m-1}\sum_{j=0}^{i}\mathcal{T}^{\textrm{(l)}}_{ij}(r)\epsilon^{2i}s^{2i-2j}w^{2j}+\sum_{i=1}^{m-1}\sum_{j=0}^{i}\mathcal{T}^{\textrm{(p)}}_{ij}(r)\epsilon^{2i}s^{2i-2j}w^{2j}+\sum_{i=1}^{m-1}\sum_{j=0}^{m-1-i}\mathcal{T}^{\textrm{(r)}}_{ij}(r)\epsilon^{2i}s^{-2j-2}w^{2i+2j+2}\nonumber\\
	&+\mathcal{O}(\epsilon^{2m}\log \epsilon).
\end{align}
\end{widetext}
The coefficients $\mathcal{D}^{(\pm)}_{ij}(r)$ are those that come from expanding the \textit{direct} part of the Hadamard parametrix $U/\sigma^{d/2-1}$ in the extended coordinates $w$ and $s$, while the coefficients $\mathcal{T}_{ij}(r)$ are those coming from the \textit{tail} part $V\,\log\sigma$. We further divide the tail coefficients into three types, $\mathcal{T}_{ij}^{\textrm{(l)}}$ which are those terms in the expansion of the tail that contain a logarithm, $\mathcal{T}_{ij}^{\textrm{(p)}}$ are those terms in the tail which are polynomial in $s^{2}$ and $w^{2}$, and $\mathcal{T}_{ij}^{(\textrm{r})}$ which are those terms in the expansion of the tail which are rational in $s^{2}$ and $w^{2}$ (unlike the polynomial terms, these are not ordinary integrable function near coincidence). There is a degeneracy in what we label as the $\mathcal{T}_{ij}^{\textrm{(p)}}$ terms since we can absorb some of the $\log(\ell^{2})\mathcal{T}_{ij}^{\textrm{(l)}}$ terms into a redefinition of the $\mathcal{T}_{ij}^{\textrm{(p)}}$ coefficients. This is simply related to the well-known ambiguity that arises due to our freedom to add factors of the homogeneous solution $V$ to the Green function, which is also crucially related to the trace anomaly. Regardless, Eq.~(\ref{eq:HadamardExp}) is our convention for how these coefficients are defined. For a massless scalar field in the $d=6$ Schwarzschild-Tangherlini spacetime, we give explicit expressions for these coefficients in Tables \ref{tab:Dcoeffplus}-\ref{tab:Tcoeff}. For higher even dimensions, the expressions are too large to be useful in print-form, however, a Mathematica Notebook containing the expressions for arbitrary metric function $f(r)$ is available online \cite{MyWebpage}.

Apart from the existence of $\log$ singularities, another important distinction in the even-dimensional case is that there are terms that are polynomial in both $\cos\kappa\Delta\tau$ and $\cos\gamma$. This is a result of the fact that only even powers of $w$ and $s$ arise in the expansion. This significantly simplifies matters since we recall that we desire mode-sum representation of the parametrix in a basis of Fourier frequency modes and Gegenbauer polynomials, and this implies, for example, that if we expand a polynomial in $\cos\gamma$ in terms of Gegenbauer polynomials $C_{l}^{\mu}(\cos\gamma)$ involves only a finite number of terms. Or put another way, there is no large $l$ contribution coming from terms that are polynomial in $\cos\gamma$. Similarly, there are no large $n$ contributions from terms that are polynomial in $\cos\kappa\Delta\tau$. This implies that it is redundant to decompose terms in the Hadamard parametrix that are polynomial in both $\cos\kappa\Delta\tau$ and $\cos\gamma$ since they cannot improve the convergence of the resultant mode-sum, since they do not contribute for large $l$ and $n$. In Eq.~(\ref{eq:HadamardExp}), terms involving coefficients $\mathcal{D}_{ij}^{(-)}$ for $j\ge d/2-1$ and terms involving the coefficients $\mathcal{T}_{ij}^{(\textrm{p})}$ are polynomial in $\cos\kappa\Delta\tau$ and $\cos\gamma$ and hence need not be expressed as a mode-sum, but rather kept in closed form. Moreover, since we are eventually interested in the coincidence limit, only the zeroth order polynomial survives, i.e., the $\mathcal{D}_{ij}^{(-)}$ term with $i=j=d/2-1$. Hence we may re-express (\ref{eq:HadamardExp}) as
\begin{widetext}
\begin{align}
	\label{eq:HadamardExpNew}
	\frac{U}{\sigma^{\frac{d}{2}-1}}+V\,\log(2\sigma/\ell^{2})=\sum_{i=0}^{ \frac{d}{2}+m-2}\sum_{j=0}^{ i}\mathcal{D}^{(+)}_{ij}(r)\epsilon^{2i-2\mu-1}\frac{w^{2i+2j}}{s^{2\mu+2j+1}}+\sum_{i=1}^{ \frac{d}{2}+m-2}\sum_{j=1}^{\min\{i,\frac{d}{2}-2\}}\mathcal{D}^{(-)}_{ij}(r)\epsilon^{2i-2\mu-1}\frac{w^{2i-2j}}{s^{2\mu-2j+1}}\nonumber\\
	+\log(\epsilon^{2}s^{2}/\ell^{2})\sum_{i=0}^{m-1}\sum_{j=0}^{i}\mathcal{T}^{\textrm{(l)}}_{ij}(r)\epsilon^{2i}s^{2i-2j}w^{2j}+\sum_{i=1}^{m-1}\sum_{j=0}^{m-1-i}\mathcal{T}^{\textrm{(r)}}_{ij}(r)\epsilon^{2i}s^{-2j-2}w^{2i+2j+2}+\mathcal{D}_{\frac{d}{2}-1,\frac{d}{2}-1}^{(-1)}(r)\nonumber\\
	+\mathcal{O}(\epsilon^{2m}\log \epsilon),
\end{align}
\end{widetext}
where we have ignored terms that are polynomial in $w^{2}$ and $s^{2}$ which vanish at coincidence. There is a slight abuse of notation here since these polynomial terms are lower order than the $\mathcal{O}(\epsilon^{2m}\log\epsilon)$ terms that we are also ignoring in the small $\epsilon$ expansion, but as explained above, the point is that the polynomial terms do not contribute to either the overall answer nor the speed of convergence. Hence they can be safely ignored.

The lowest order truncation of this series that captures all the singular terms is for $m=0$, however, it is generally useful to keep higher-order terms which can greatly improve the convergence properties of the mode-sum expression for the regularized Green function. More specifically, if we subtract only the $m=0$ terms from the Green function (in a procedure explained in the next section), then the resultant regularized Green function is convergent, since we have captured all the divergences, but in the coincidence limit the convergence is only conditional. This implies that the order in which the mode-sum is performed matters and is actually tied to the order in which we take coincidence limits (see Ref.~\cite{OttewillTaylorCS} for a discussion of this point). In the present context, the order in which the limits are taken is already fixed by our choice of $w$ and $s$, since it is not possible to take $s\to 0$ without first taking $w\to 0$. This implies that the $n$-sum ought to be summed first to get the correct answer. In the standard coordinate approach, one can take $\Delta \tau\to 0$ independently from $\gamma\to 0$ and whichever limit is taken first, the corresponding sum ought to be performed first. For example, if we take $\Delta\tau\to 0$ first, then the Fourier frequency sum is the inner sum of the resultant mode-sum expression. Returning to our approach, if we further include the $m=1$ terms then all the terms we are ignoring formally vanish in the coincidence limit. However, the convergence of the mode-sum is still relatively slow and a high-accuracy calculation can be computationally expensive. It is therefore prudent to include still higher-order terms in the Hadamard parametrix, even though these terms formally vanish, their inclusion serves to speed the convergence of the mode-sum, at least in principle. In practice however, including very high-order terms becomes prohibitively slow simply because there are so many terms to compute in the decomposition of the Hadamard parametrix. We find that including terms up to $m=2$ or $m=3$ is the most computationally efficient. In what follows, we take $m=2$ for concreteness, which gives
\begin{widetext}
\begin{align}
	\label{eq:Hadamardclosed}
	\frac{U}{\sigma^{\frac{d}{2}-1}}+V\,\log(2\sigma/\ell^{2})&=\sum_{i=0}^{ \frac{d}{2}}\sum_{j=0}^{ i}\mathcal{D}^{(+)}_{ij}(r)\epsilon^{2i-2\mu-1}\frac{w^{2i+2j}}{s^{2\mu+2j+1}}+\sum_{i=1}^{ \frac{d}{2}}\sum_{j=1}^{ \min\{i,\frac{d}{2}-2\}}\mathcal{D}^{(-)}_{ij}(r)\epsilon^{2i-2\mu-1}\frac{w^{2i-2j}}{s^{2\mu-2j+1}}\nonumber\\
	&+\log(\epsilon^{2}s^{2}/\ell^{2})\sum_{i=0}^{1}\sum_{j=0}^{i}\mathcal{T}^{\textrm{(l)}}_{ij}(r)\epsilon^{2i}s^{2i-2j}w^{2j}+\mathcal{T}^{\textrm{(r)}}_{10}(r)\epsilon^{2i}s^{-2}w^{4}+\mathcal{D}_{\frac{d}{2}-1,\frac{d}{2}-1}^{(-)}(r)+\mathcal{O}(\epsilon^{4}\log \epsilon),
\end{align}
\end{widetext}
where again we have ignored the terms that are polynomial in $w^{2}$ and $s^{2}$. An important point to emphasize here is that we have maintained a point-splitting in multiple directions in this expression. It is, of course, tempting to simplify this expression by choosing only one direction in which to point split and if this direction was a Killing direction, then the resultant parametrix would be very simple indeed. This is the procedure that practically all other regularization schemes adopt. Our perspective, which is a major departure from the standard one, is that employing this freedom to point split in any direction too early in the calculation actually hinders rather than helps. Recall that what is actually needed is a mode-by-mode subtraction of the physical Green function (representing the propagation of a quantum field in some quantum state) minus the Hadamard parametrix (which captures the local ultra-violet behaviour of the Green function) and this mode-by-mode subtraction arises most naturally if we split in multiple directions. This will be shown in the next section where an explicit mode-sum decomposition for the Hadamard parametrix is derived.

\section{Mode-Sum Representation of the Hadamard Parametrix}
We wish to decompose the terms of the Hadamard parametrix (\ref{eq:Hadamardclosed}) in terms of Fourier frequency modes and multipole moments. If this can be achieved then a mode-by-mode subtraction for the regularized Green function is feasible. We will consider the direct and tail parts separately.

\subsection{Regularization Parameters for the Direct Part}
In this section, we decompose terms of the form $w^{2i\pm 2j}/s^{2\mu\pm 2j+1}$ and invert to compute the regularization parameters. The derivation here is very similar to that needed to compute regularization parameters in arbitrary odd dimensions \cite{TaylorBreenOdd}, except that the parameter $\mu=(d-3)/2$ is now half-integer. As in \cite{TaylorBreenOdd}, the starting point is to assume a Fourier frequency and multipole decomposition of the form
\begin{align}
	\frac{w^{2i\pm 2j}}{s^{2\mu\pm 2j+1}}=\sum_{n=-\infty}^{\infty}e^{in\kappa\Delta\tau}\sum_{l=0}^{\infty}(2 l+2\mu)C_{l}^{\mu}(\cos\gamma)\nonumber\\
	\times\,\,\stackrel{[d]}{\Psi}\!\!{}_{nl}^{(\pm)}(i,j|r),
\end{align}
which we try to invert to determine the regularization parameters $\stackrel{[d]}{\Psi}\!\!{}_{nl}^{(\pm)}(i, j|r)$. With $x=\cos\gamma$, multiplying both sides by $e^{-i n'\Delta \tau}(1-x^{2})^{\mu-\frac{1}{2}}C_{l'}^{\mu}(x)$ and integrating gives
\begin{align}
	\stackrel{[d]}{\Psi}\!\!{}_{nl}^{(\pm)}(i, j|r)=\frac{\kappa}{(2\pi)^{2}}\frac{2^{2\mu-1}\Gamma(\mu)^{2}l!}{\Gamma(l+2\mu)}\int_{0}^{2\pi/\kappa}\int_{-1}^{1}\frac{w^{2i\pm 2j}}{s^{2\mu\pm 2j+1}}\nonumber\\
	\times\,\,e^{-i n \kappa\Delta\tau}(1-x^{2})^{\mu-\frac{1}{2}}C_{l}^{\mu}(x)\,dx\,d\Delta\tau,
\end{align}
where we have used the completeness relations
\begin{align}
	\label{eq:completeness}
	&\int_{0}^{2\pi/\kappa}e^{-i (n-n')\Delta\tau}d\Delta\tau=\frac{2\pi}{\kappa}\delta_{n n'},\nonumber\\
	&\int_{-1}^{1}(1-x^{2})^{\mu-\frac{1}{2}}C_{l}^{\mu}(x)C_{l'}^{\mu}(x)\,dx=\frac{2^{1-2\mu}\pi\,\Gamma(n+2\mu)}{(l+\mu)\,l!\,\Gamma(\mu)^{2}}\delta_{l l'}.
\end{align}
We perform the $x$ integration above by employing the identity \cite{CohlGegenInt}
\begin{align}
	\label{eq:CohlInt}
&\int_{-1}^{1}\frac{(1-x^{2})^{\mu-1/2}C_{l}^{\mu}(x)}{(z-x)^{\mu\pm j+1/2}}dx\nonumber\\
&=\frac{(-1)^{j}\sqrt{\pi}\Gamma(l+2\mu)(z^{2}-1)^{\mp j/2}}{2^{\mu-3/2}l!\Gamma(\mu)\Gamma(\mu\pm j+1/2)}Q^{\pm j}_{l+\mu-1/2}(z),	
\end{align}
to obtain
\begin{align}
	\label{eq:RegParamInt}
	&\stackrel{[d]}{\Psi}\!\!{}_{nl}^{(\pm)}(i,  j|r)=\frac{\kappa}{(2\pi)^{2}}\frac{2^{i}\sqrt{\pi}(-1)^{j}\Gamma(\mu)}{\kappa^{2i \pm 2j}r^{2\mu\pm 2j+1}\Gamma(\mu+\frac{1}{2}\pm j)}\nonumber\\
	&\times\,\int_{0}^{2\pi/\kappa}(1-\cos\kappa t)^{i\pm j}e^{-i n \kappa t}(z^{2}-1)^{\mp j/2}Q^{\pm  j}_{l+\mu-\frac{1}{2}}(z) dt,
\end{align}
with
\begin{align}
	\label{eq:zDef}
	z=1+\frac{f^{2}}{\kappa^{2}r^{2}}(1-\cos\kappa t).
\end{align}
As already mentioned, for even $d\ge 4$, the parameter $\mu=(d-3)/2$ is always half-integer, implying that $\mu-\tfrac{1}{2}\in \mathbb{N}$. In particular, we note that since $l+\mu-\tfrac{1}{2}+j$ is a positive integer, the associated Legendre function of the second kind appearing in the integral representation of the regularization parameters above is always well-defined. Also, for the $\mathcal{D}_{ij}^{(-)}(r)$ terms, the largest value that $j$ can assume is $d/2-2$ and hence $l+\mu-\tfrac{1}{2}-j$ cannot be a negative integer and the right-hand side of (\ref{eq:RegParamInt}) remains meaningful. We focus first on the former case of computing the $\stackrel{[d]}{\Psi}\!\!{}_{nl}^{(+)}(i, j|r)$ terms. Using the fact that
\begin{align}
	\label{eq:LegDeriv}
	(z^{2}-1)^{-j/2}Q_{\nu}^{j}(z)=\frac{(-1)^{j}}{2^{j}(1-\cos\kappa t)^{j}}\Big(\frac{1}{\eta}\frac{\partial}{\partial\eta}\Big)^{j}Q_{\nu}(z),
\end{align}
where
\begin{align}
	\eta\equiv \sqrt{1+\frac{f(r)}{\kappa^{2}r^{2}}},
\end{align}
we arrive at
\begin{align}
	&\stackrel{[d]}{\Psi}\!\!{}_{nl}^{(+)}(i,j|r)=\frac{\kappa}{(2\pi)^{2}}\frac{2^{i-j}\sqrt{\pi}\Gamma(\mu)}{\kappa^{2i \pm 2j}r^{2\mu+ 2j+1}\Gamma(\mu+\frac{1}{2}+j)}\nonumber\\
	&\times\,\Bigg(\frac{1}{\eta}\frac{\partial}{\partial\eta}\Bigg)^{j}\int_{0}^{2\pi/\kappa}(1-\cos\kappa t)^{i}e^{-i n \kappa t}Q_{l+\mu-\frac{1}{2}}(z) dt.
\end{align}
In order to perform the integral we must factor out the time dependence from the Legendre function, which may be achieved by employing the addition theorem \cite{GradRiz},
\begin{align}
	\label{eq:AddThm}
	Q_{\nu}(z)=P_{\nu}(\eta)Q_{\nu}(\eta)+2\sum_{p=1}^{\infty}(-1)^{p}P_{\nu}^{-p}(\eta)Q_{\nu}^{p}(\eta)\cos p\,\kappa t,
\end{align}
whence the time integral reduces to
\begin{align}
	\label{eq:timeint}
	&\int_{0}^{2\pi/\kappa}(1-\cos\kappa t)^{i}e^{-i n\kappa t}\cos p \kappa t\,dt=\nonumber\\
	&\frac{\sqrt{\pi }}{\kappa}\Bigg[ \frac{2^i i! \Gamma \left(i+\frac{1}{2}\right) (-1)^{n-p}}{(i+p-n)! (i-p+n)!}+\frac{2^i
   i! \Gamma \left(i+\frac{1}{2}\right) (-1)^{p+n}}{(i-p-n)! (i+p+n)!}\Bigg].
\end{align}
The factorials in the denominator imply that there is a finite number of integer $p$ for which the integral is nonzero. In particular, the first term on the right-hand side of (\ref{eq:timeint}) is nonzero only for $|p-n|\le i$ while the second term is nonzero for $|p+n|\le i$. The range is further restricted in our case since $p\ge 1$ and hence the sets of integers $p$ for which the first and second terms are nonzero are $p\in \{\max(1,n-i),n+i\}$ and $p\in\{\max(1, -n-i),i-n\}$, respectively. An equivalent expression for (\ref{eq:timeint}) in terms of a sum of Kronecker deltas is easily derived. Putting these together, we obtain
\begin{widetext}
\begin{align}
	\label{eq:RegParamP}
\stackrel{[d]}{\Psi}\!\!{}_{nl}^{(+)}(i,j|r)&=\frac{2^{2i-j-1}(-1)^{n}i!\,\Gamma(i+\frac{1}{2})\Gamma(\mu)}{\pi\kappa^{2i + 2j}r^{2\mu+ 2j+1}\Gamma(j+\mu+\tfrac{1}{2})}\left(\frac{1}{\eta}\frac{d}{d \eta}\right)^{j}\Bigg\{\frac{P_{l+\mu-\frac{1}{2}}(\eta)Q_{l+\mu-\frac{1}{2}}(\eta)}{(i-n)!(i+n)!}\nonumber\\
&+\sum_{p=\max\{1,n-i\}}^{i+n}\frac{P^{-p}_{l+\mu-\frac{1}{2}}(\eta)Q^{p}_{l+\mu-\frac{1}{2}}(\eta)}{(i+p-n)!(i-p+n)!}+\sum_{p=\max\{1,-n-i\}}^{i-n}\frac{P^{-p}_{l+\mu-\frac{1}{2}}(\eta)Q^{p}_{l+\mu-\frac{1}{2}}(\eta)}{(i+p+n)!(i-p-n)!}\Bigg\}.
\end{align}

We turn now to the $\stackrel{[d]}{\Psi}\!\!{}_{nl}^{(-)}(i, j|r)$ terms. Again, the derivation is analogous to the odd-dimensional case presented in Ref.~\cite{TaylorBreenOdd} so we omit much of the details. Briefly,  making use of the identity
\begin{align}
	\label{eq:GenAddition}
(z^{2}-1)^{j/2}Q^{-j}_{\nu}(z)=\sum_{k=0}^{j}\frac{(-1)^{k}}{2^{j+1}}\binom{j}{k}\frac{(2\nu+2 j-4k+1)}{(\nu-k+\frac{1}{2})_{j+1}}Q_{\nu+j-2k}(z)
\end{align}
in (\ref{eq:RegParamInt}) gives
\begin{align}
	\stackrel{[d]}{\Psi}\!\!{}_{nl}^{(-)}(i,  j|r)=\frac{\kappa}{(2\pi)^{2}}\frac{2^{i-j}\sqrt{\pi}(-1)^{j}\Gamma(\mu)}{\kappa^{2i - 2j}r^{2\mu- 2j+1}\Gamma(\mu+\frac{1}{2}- j)}\sum_{k=0}^{j}(-1)^{k}\binom{j}{k}\frac{(l+\mu+j-2k)}{(l+\mu-k)_{j+1}}\nonumber\\
	\times\,\int_{0}^{2\pi/\kappa}(1-\cos\kappa t)^{i- j}e^{-i n \kappa t}Q_{l+\mu-\frac{1}{2}+j-2k}(z) dt.
\end{align}
We now proceed as above: we apply the addition theorem (\ref{eq:AddThm}) to isolate the time-dependence, and integrate using (\ref{eq:timeint}). The result is
\begin{align}
	\label{eq:RegParamN}
\stackrel{[d]}{\Psi}\!\!{}_{nl}^{(-)}(i,j|r)&=\frac{2^{2i-2j-1}(-1)^{n+j}(i-j)!\Gamma(i-j+\tfrac{1}{2})\Gamma(\mu)}{\pi\kappa^{2i - 2j}r^{2\mu- 2j+1}\Gamma(\mu+\tfrac{1}{2}-j)}\sum_{k=0}^{j}(-1)^{k}\binom{j}{k}\frac{(l+\mu+j-2k)}{(l+\mu-k)_{j+1}}\nonumber\\
&\times\,\,\Bigg\{\frac{P_{l+\mu-\frac{1}{2}+j-2k}(\eta)Q_{l+\mu-\frac{1}{2}+j-2k}(\eta)}{(i-j-n)!(i-j+n)!}+\sum_{p=\max\{1,n-i+j\}}^{i-j+n}\frac{P^{-p}_{l+\mu-\frac{1}{2}+j-2k}(\eta)Q^{p}_{l+\mu-\frac{1}{2}+j-2k}(\eta)}{(i-j+p-n)!(i-j-p+n)!}\nonumber\\
&+\sum_{p=\max\{1,-n-i+j\}}^{i-j-n}\frac{P^{-p}_{l+\mu-\frac{1}{2}+j-2k}(\eta)Q^{p}_{l+\mu-\frac{1}{2}+j-2k}(\eta)}{(i-j+p+n)!(i-j-p-n)!}\Bigg\}.
\end{align}
\end{widetext}

Eqs. (\ref{eq:RegParamP}) and (\ref{eq:RegParamN}) are the regularization parameters for the direct part of the scalar two-point function in a static spherically symmetric spacetime in arbitrary even dimensions.

\subsection{Regularization Parameters for the Tail}
In this section, we derive regularization parameters for the tail terms in the Hadamard parametrix. Recall that we divided the tail term into subcategories depending on whether there was a logarithm, whether the terms were polynomial in $s^{2}$ and $w^{2}$ or whether they were rational (generalized) functions of $w^{2}$ and $s^{2}$. We have already explained that the polynomial terms do not need to be expressed as a mode-sum, even if they were, the sums would be finite and hence offer no advantage in improving the convergence of the mode-sum expression for the vacuum polarization. The only rational term coming from the tail term in the parametrix at the order being considered is a term of the form $w^{4}/s^{2}$. This type of term has in fact already been considered since it shows up in the direct part corresponding to the coefficient $\mathcal{D}_{ij}^{(-)}(r)$ with $i=\frac{d}{2}$, $j=\frac{d}{2}-2$. Thus, the corresponding regularization parameter is given by (\ref{eq:RegParamN}) with $i=\frac{d}{2}$, $j=\frac{d}{2}-2$. Equivalently, we could simply absorb the term $\mathcal{T}_{10}^{(\textrm{r})}(r)$ into the direct part in (\ref{eq:Hadamardclosed}) by the redefinition
\begin{align}
	\mathcal{\tilde{D}}^{(-)}_{\frac{d}{2},\frac{d}{2}-2}(r)=\mathcal{D}^{(-)}_{\frac{d}{2},\frac{d}{2}-2}(r)+\mathcal{T}_{10}^{(\textrm{r})}(r).
\end{align}

Hence, the only remaining terms in the tail that need to be considered are those that involve a log, i.e., those of the form $\log(s^{2}/\ell^{2})\,s^{2i-2j}w^{2j}$. As before the starting point is to assume an expansion of the form
\begin{align}
	s^{2i-2j}w^{2j}\log\left(\frac{s^{2}}{\ell^{2}}\right)=\sum_{n=-\infty}^{\infty}e^{in\kappa\,\Delta\tau}\sum_{l=0}^{\infty}(2l+2\mu)C_{l}^{\mu}(\cos\gamma)\nonumber\\
	\times\,\,\stackrel{[d]}{\chi}\!\!{}_{nl}(i,j|r).
\end{align}
Using the completeness relations (\ref{eq:completeness}), we can invert to get the double integral representation for the regularization parameters:
\begin{align}
	\label{eq:chiInt}
	\stackrel{[d]}{\chi}\!\!{}_{nl}(i,j|r)=\frac{\kappa}{(2\pi)^{2}}\frac{2^{2\mu-1}\Gamma(\mu)^{2}l!}{\Gamma(l+2\mu)}\int_{0}^{2\pi/\kappa}\int_{-1}^{1}w^{2j}\,e^{-i n \kappa\Delta\tau}\nonumber\\
	\times\,\log(s^{2}/\ell^{2})s^{2i-2j}(1-x^{2})^{\mu-\frac{1}{2}}C_{l}^{\mu}(x)\,dx\,d\Delta\tau.
\end{align}
For the angular integral, we wish to find a useful expression for the integrals of the form
\begin{align}
	\label{eq:LogParamInt}
	\int_{-1}^{1}\log(z-x)\,(z-x)^{k}(1-x^{2})^{\mu-\frac{1}{2}}C_{l}^{\mu}(x)\,dx,
\end{align}
where $k=i-j\in \mathbb{N}$ and we remind the reader that $z>1$ is given by Eq.~(\ref{eq:zDef}). 
\begin{widetext}

There are several ways one might proceed, for example, one could attempt to differentiate $(z-x)^{\lambda}$ with respect to the exponent and then take the limit $\lambda\to i-j$. This indeed can be done but is needlessly complicated for all but a handful of low $l$-modes. Instead, we start with the Rodrigues'-type formula for the Gegenbauer polynomials \cite{GradRiz}
\begin{align}
	C_{l}^{\mu}(x)=\frac{(-1)^{l}(2\mu)_{l}}{2^{l}l!(\mu+\frac{1}{2})_{l}}(1-x^{2})^{\frac{1}{2}-\mu}\frac{d^{l}}{dx^{l}}\Big[(1-x^{2})^{\mu+l-\frac{1}{2}}\Big].
\end{align} 
Substituting this into Eq.~(\ref{eq:LogParamInt}) and integrating by parts $l$ times gives
\begin{align}
	\label{eq:LogInt}
\int_{-1}^{1}\log(z-x)\,(z-x)^{k}(1-x^{2})^{\mu-\frac{1}{2}}C_{l}^{\mu}(x)\,dx=\frac{(2\mu)_{l}}{2^{l}l!(\mu+\frac{1}{2})_{l}}\int_{-1}^{1}B_{lk}(z,x)(1-x^{2})^{l+\mu-\frac{1}{2}}dx,
\end{align}
where
\begin{equation}
	\label{eq:Bcoeff}
	B_{lk}(z,x)=\begin{cases}
	(-1)^{k+1}k!(l-k-1)!(z-x)^{k-l} & l>k\\ \\
	(-1)^{l}(k-l+1)_{l}(z-x)^{k-l}\big\{\log(z-x)+\psi(k+1)-\psi(k+1-l)\big\} & l\le k.
	 \end{cases}
\end{equation}
We note that in our case, $k=i-j$ will be a small number and hence all but the lowest-lying $l$-modes will satisfy $l>i-j$. In fact, to compute the regularized Green function up to $\mathcal{O}(\epsilon^{4}\log\epsilon)$ requires only the $k=0,1$ modes and hence all $l\ge 2$ will be given by the more simple expression for $B_{lk}$ above. In this case, the integral in Eq.~(\ref{eq:LogInt}) is expressible in terms of the Legendre function of the second kind,
\begin{align}
	\int_{-1}^{1}\log(z-x)\,(z-x)^{k}(1-x^{2})^{\mu-\frac{1}{2}}C_{l}^{\mu}(x)\,dx=(-1)^{\mu-\frac{1}{2}}2^{\mu+\frac{1}{2}}\frac{k!}{l!}(2\mu)_{l}\Gamma(\mu+\frac{1}{2})(z^{2}-1)^{\frac{1}{2}(\mu+k+\frac{1}{2})}Q_{l+\mu-\frac{1}{2}}^{-\mu-k-\frac{1}{2}}(z),\nonumber\\
	\textrm{for}\quad l>k.
\end{align}
% {\color{red}There's a typo in Eq. 8.712 in Gradshtyn Rhyzik, the factor $(z^{2}-1)^{-\mu/2}$ should be $(z^{2}-1)^{\mu/2}$ and the restriction to $\textrm{Re}(\mu)>-1$ doesn't seem to be necessary!}
Hence, substituting this result into (\ref{eq:chiInt}) and using some standard identities involving Gamma functions gives
\begin{align}
	\stackrel{[d]}{\chi}\!\!{}_{nl}(i,j|r)=\frac{\kappa}{(2\pi)^{2}}(-1)^{\mu-\frac{1}{2}}(i-j)!(2r^{2})^{i-j}\sqrt{\pi}2^{\mu+\frac{1}{2}}\Gamma(\mu)\int_{0}^{2\pi/\kappa}w^{2j}\,e^{-i n \kappa\Delta\tau}(z^{2}-1)^{\frac{1}{2}(\mu+\frac{1}{2}+i-j)}Q_{l+\mu-\frac{1}{2}}^{-\mu-i+j-\frac{1}{2}}(z)d\Delta\tau\nonumber\\
	\textrm{for}\quad l>i-j.
\end{align}
Note that the terms involving the arbitrary length-scale $\log(\ell^{2})$ vanishes for $l>i-j$. To perform the time integral, we again adopt Eqs. (\ref{eq:GenAddition}) and (\ref{eq:AddThm}) to factor out the time dependence in terms of exponentials, in much the same way we did for the regularization parameters of the direct part. The result for the regularization parameters for the $\log$ terms with $l>i-j$ is
\begin{align}
	\label{eq:RegParamLogLarge}
\stackrel{[d]}{\chi}\!\!{}_{nl}(i,j|r)=(-1)^{n+\mu-\frac{1}{2}}\frac{2^{2j-1}r^{2i-2j}\Gamma(\mu)(i-j)!j!\Gamma(j+\frac{1}{2})}{\pi\,\kappa^{2j}}\sum_{k=0}^{\mu+\frac{1}{2}+i-j}(-1)^{k}\binom{\mu+\frac{1}{2}+i-j}{k}\frac{(l+2\mu+i-j-2k+\frac{1}{2})}{(l+\mu-k)_{\mu+\frac{3}{2}+i-j}}\nonumber\\
\times\Bigg\{\frac{P_{l+2\mu+i-j-2k}(\eta)\,Q_{l+2\mu+i-j-2k}(\eta)}{(j-n)!(j+n)!}+\sum_{p=\max\{1,n-j\}}^{j+n}\frac{P_{l+2\mu+i-j-2k}^{-p}(\eta)\,Q^{p}_{l+2\mu+i-j-2k}(\eta)}{(j+p-n)!(j-p+n)!}\nonumber\\
+\sum_{p=\max\{1,-n-j\}}^{j-n}\frac{P_{l+2\mu+i-j-2k}^{-p}(\eta)\,Q^{p}_{l+2\mu+i-j-2k}(\eta)}{(j+p+n)!(j-p-n)!}\Bigg\},\nonumber\\
\textrm{for}\quad l>i-j.
\end{align}

For the $l\le i-j$ modes, we have not found it most convenient to proceed as above, mainly because the integral (\ref{eq:LogInt}) is not easily performed in terms of known functions, let alone performing the additional time integral in Eq.~(\ref{eq:chiInt}). Instead, we rewrite (\ref{eq:chiInt}) succinctly by adopting Eqs.~(\ref{eq:LogInt})-(\ref{eq:Bcoeff}) and expressing in terms of a derivative with respect to the exponent,
\begin{align}
	\label{eq:RegParamLogDoubleInt}
	\stackrel{[d]}{\chi}\!\!{}_{nl}(i,j|r)=\frac{\kappa}{(2\pi)^{2}}\frac{\sqrt{\pi}\Gamma(\mu)(2r^{2})^{i-j}(-1)^{l}}{2^{l}\Gamma(\mu+l+\frac{1}{2})}\Big(\frac{2}{\kappa^{2}}\Big)^{j}\Bigg[\frac{d}{d\lambda}\int_{0}^{2\pi/\kappa}(1-\cos\kappa t)^{j}e^{-in\kappa t}\nonumber\\
	\times\,\,\int_{-1}^{1}(1-x^{2})^{l+\mu-\frac{1}{2}}(\lambda+1-l)_{l}(z-x)^{\lambda-l}\Big(\frac{2r^{2}}{\ell^{2}}\Big)^{\lambda-i+j}dx\,dt\Bigg]_{\lambda=i-j}.
\end{align}
The $x$-integral here can be performed in terms of Olver's definition \cite{Olver} of the associated Legendre function of the second kind. Unlike the usual definition of the Legendre function, Olver's has the advantage that it is valid for all values of the parameters. In particular, we have
\begin{align}
	\label{eq:intOlverQ}
	\int_{-1}^{1}(1-x^{2})^{l+\mu-\frac{1}{2}}(z-x)^{\lambda-l}dx=(z^{2}-1)^{\frac{1}{2}(\mu+\lambda+\frac{1}{2})}2^{l+\mu+\frac{1}{2}}\Gamma(l+\mu+\frac{1}{2})\mathcal{Q}_{l+\mu-\frac{1}{2}}^{-\mu-\lambda-\frac{1}{2}}(z),
\end{align}
where $\mathcal{Q}_{\nu}^{\mu}(z)$ is Olver's Legendre function of the second kind defined by the Hypergeometric series
\begin{align}
	\mathbf{\mathcal{Q}}^{\mu}_{\nu}(z)=\frac{\sqrt{\pi}}{2^{\nu+1}\,z^{\mu+\nu+1}}(z^{2}-1)^{\mu/2}\mathbf{F}\left(\tfrac{1}{2}\mu+\tfrac{1}{2}\nu+\tfrac{1}{2},\tfrac{1}{2}\mu+\tfrac{1}{2}\nu+1;\nu+\tfrac{3}{2};\tfrac{1}{z^{2}}\right)
\end{align}
with $\mathbf{F}(a,b;c;z)$ the regularized Hypergeometric function. Whenever $\mu+\nu$ is not a negative integer, Olver's definition is related to the standard one by
\begin{align}
	\mathbf{\mathcal{Q}}^{\mu}_{\nu}(z)=\frac{e^{-\mu\,\pi\,i}Q^{\mu}_{\nu}(z)}{\Gamma(\mu+\nu+1)}.
\end{align}
Employing (\ref{eq:intOlverQ}) in (\ref{eq:RegParamLogDoubleInt}) results in
\begin{align}
	\label{eq:RegParamLogInt}
	\stackrel{[d]}{\chi}\!\!{}_{nl}(i,j|r)=\frac{\kappa}{(2\pi)^{2}}\sqrt{\pi}2^{\mu+\frac{1}{2}}\Gamma(\mu)(2r^{2})^{i-j}(-1)^{l}\Big(\frac{2}{\kappa^{2}}\Big)^{j}\Bigg[\frac{d}{d\lambda}(\lambda+1-l)_{l}\Big(\frac{2r^{2}}{\ell^{2}}\Big)^{\lambda-i+j}\int_{0}^{2\pi/\kappa}(1-\cos\kappa t)^{j}e^{-in\kappa t}\nonumber\\
	\times\,\,(z^{2}-1)^{\frac{1}{2}(\mu+\lambda+\frac{1}{2})}\mathcal{Q}_{l+\mu-\frac{1}{2}}^{-\mu-\lambda-\frac{1}{2}}(z)dt\Bigg]_{\lambda=i-j}\qquad \textrm{for}\quad l\le i-j.
\end{align}
Now, one can in principle, perform the derivative with respect to $\lambda$ (see, for example, \cite{CohlDerivQ} for derivatives of associated Legendre functions with respect to the order) and subsequently perform the integral here to obtain series expansions in terms of products of Legendre functions. The resultant expressions are extremely complicated so we take the more pragmatic approach of simply performing the integral above numerically, which presents no difficulty. Moreover, these numerical integrals are only needed for $l=0,1$ at the order being considered, and hence there is no loss of efficiency in the calculation as a result of not using a closed-form representation.

% Now, using the binomial theorem, we have
% \begin{align}
% 	(z-x)^{\lambda-l}=\sum_{p=0}^{\infty}\sum_{q=0}^{\infty}\frac{(-1)^{p}}{p!q!}\frac{\Gamma(\lambda-l+1)}{\Gamma(\lambda-l-p-q+1)}2^{\lambda-l-p-q}(z-1)^{q}(1-x)^{p}.
% \end{align}
% Substituting this in above and performing the standard integrals gives
% \begin{align}
% 	\stackrel{[d]}{\chi}\!\!{}_{nl}(i,j|r)=\frac{\Gamma(\mu)r^{2i-2j}2^{2i+2\mu-1}(-1)^{l+n}}{\pi\,\kappa^{2j}}\Bigg[\frac{\partial}{\partial\lambda}\Big(\frac{4 r^{2}}{\ell^{2}}\Big)^{\lambda-i+j}\sum_{p=0}^{\infty}\sum_{q=0}^{\infty}\frac{(-1)^{p}}{p!q!}\frac{\Gamma(\lambda+1)}{\Gamma(\lambda-l-p-q+1)}\Big(\frac{f}{\kappa^{2}r^{2}}\Big)^{q}\nonumber\\
% 	\times \frac{(j+q)!\Gamma(j+q+\frac{1}{2})}{(j+q-n)!(j+q+n)!}\frac{\Gamma(l+\mu+p+\frac{1}{2})}{\Gamma(2l+2\mu+p+1)}\Bigg]_{\lambda=i-j}
% \end{align}
\end{widetext}

\subsection{The Hadamard Parametrix}
Combining the results of the previous subsections, the mode-sum representation of the Hadamard parametrix for a scalar field in a static spherically symmetric even-dimensional spacetime is (ignoring terms that do not contribute in the coincidence limit and henceforth setting the book-keeping parameter $\epsilon$ to unity)
\begin{align}
	G_{\subS}&(x,x')=\frac{\Gamma(\frac{d}{2}-1)}{2(2\pi)^{d/2}}\sum_{l=0}^{\infty}(2l+2\mu)C_{l}^{\mu}(\cos\gamma)\sum_{n=-\infty}^{\infty}e^{in\kappa\Delta\tau}\nonumber\\
	&\times\,\,\Bigg\{\sum_{i=0}^{\frac{d}{2}}\sum_{j=0}^{i}\mathcal{D}_{ij}^{(+)}(r)\stackrel{[d]}{\Psi}\!\!{}_{nl}^{(+)}(i,j|r)\nonumber\\
	&+\sum_{i=1}^{\frac{d}{2}}\sum_{j=1}^{\min\{i,\frac{d}{2}-2\}}\mathcal{D}_{i,j}^{(-)}(r)\stackrel{[d]}{\Psi}\!\!{}_{nl}^{(-)}(i,j|r)\nonumber\\
	&+\mathcal{T}_{10}^{(\textrm{r})}(r)\,\stackrel{[d]}{\Psi}\!\!{}_{nl}^{(-)}(\tfrac{d}{2},\tfrac{d}{2}-2|r)+\sum_{i=0}^{2}\sum_{j=0}^{i}\mathcal{T}_{ij}^{\textrm{(l)}}(r)\,\stackrel{[d]}{\chi}\!\!{}_{nl}(i,j|r)\Bigg\}\nonumber\\
	&+\frac{\Gamma(\frac{d}{2}-1)}{2(2\pi)^{d/2}}\mathcal{D}_{\frac{d}{2}-1,\frac{d}{2}-1}^{(-)}(r),
\end{align}
where $\stackrel{[d]}{\Psi}\!\!{}_{nl}^{(\pm)}(i,j|r)$, $\stackrel{[d]}{\chi}\!\!{}_{nl}(i,j|r)$ are given by Eqs.~(\ref{eq:RegParamP}), (\ref{eq:RegParamN}), (\ref{eq:RegParamLogLarge}) and (\ref{eq:RegParamLogInt}). This is the main result. It allows one to numerically compute the regularized vacuum polarization in arbitrary even dimensions in an extremely efficient way. We describe this calculation for a massless scalar field in the Schwarzschild-Tangherlini spacetimes in the following section.

\section{Vacuum Polarization in Schwarzschild-Tangherlini Spacetimes}
In this section we outline the numerical implementation of the regularization scheme described above to the calculation of the vacuum polarization for a scalar field in the Hartle-Hawking state on the background of even-dimensional Schwarzschild-Tangherlini black hole spacetimes. In Schwarzschild coordinates, the Schwarzschild-Tangherlini metric takes the form (\ref{eq:metric}) with
\begin{align}
	\label{eq:fST}
	f(r)=1-\Big(\frac{r_{\textrm{h}}}{r}\Big)^{d-3}.
\end{align}
These coordinates are singular at $r=r_{\textrm{h}}$ which corresponds to the black hole horizon. For simplicity, throughout the remainder of this section, we work in units where $r_{\textrm{h}}=1$, so that all lengths having numerical values are in units of the event horizon radius. That implies, for example, that the surface gravity $\kappa=\tfrac{1}{2}f'(r_{\textrm{h}})=\tfrac{1}{2}(d-3)=\mu$. We also choose the lengthscale $\ell$ in the Hadamard parametrix to satisfy $\ell=1$. Now, the regularized vacuum polarization for a scalar field in the Hartle-Hawking state is defined by the coincidence limit of the difference between the Euclidean Green function and the Hadamard parametrix,
\begin{align}
	\label{eq:phi2Def}
\langle \phi^{2} \rangle_{\textrm{reg}}&=\lim_{x'\to x}\Big\{G(x,x')-G_{\subS}(x,x')\Big\}.
\end{align}
It is clear that to compute the vacuum polarization in this context requires two ingredients: first, a mode-sum expression for the Euclidean Green function which includes accurate numerical data for the radial Green function, see Eqs.~(\ref{eq:modeGreen})-(\ref{eq:radialgreenfn}), and second, a mode-sum representation for the Hadamard parametrix which we have derived in detail in the previous section.

\subsection{Mode-Sum Calculation}
As mentioned, in order to calculate the vacuum polarization we must first calculate the radial Green function (\ref{eq:radialgreenfn}) which is a normalized product of homogeneous solutions $p_{nl}(r)$ and $q_{nl}(r)$ to Eq.~(\ref{eq:radialeqn}) satisfying boundary conditions of regularity on the horizon and at infinity, respectively. For $f(r)$ given by (\ref{eq:fST}), solutions cannot in general be given in terms of known functions and must be solved numerically, with the exception of the zero frequency modes which are
\begin{align}
\label{eqn:n0}
p_{0l}(r)&=P_{l/(d-3)} (2r^{d-3}-1)\nonumber\\
q_{0l}(r)&=Q_{l/(d-3)} (2r^{d-3}-1),
\end{align}
where $P_\nu(z)$ and $Q_\nu(z)$ are Legendre functions of the first and second kind, respectively. For the remaining modes, their numerical calculation is outlined in detail in the companion paper \cite{TaylorBreenOdd}, so the details will be omitted here.

Equipped with both accurate numerical evaluation of the radial modes for the Euclidean Green function and explicit closed-form expressions for the regularization parameters, we are now in a position to calculate the vacuum polarization $\langle \phi^2 \rangle_{\textrm{reg}}$ in even-dimensional Schwarzschild-Tangherlini spacetimes. We present results for $d=4, 6, 8, 10$. Though results for $d=4$ are long-established \cite{CandelasHowardPhi2}, we include our results here as a non-trivial check against the tabulated data in \cite{CandelasHowardPhi2}. For the higher even dimensions, to the best of our knowledge, these are the first exact results for vacuum polarization on the entire exterior Schwarzschild-Tangherlini geometries. 

\begin{figure}[ht!]
\includegraphics[width=0.95\columnwidth]{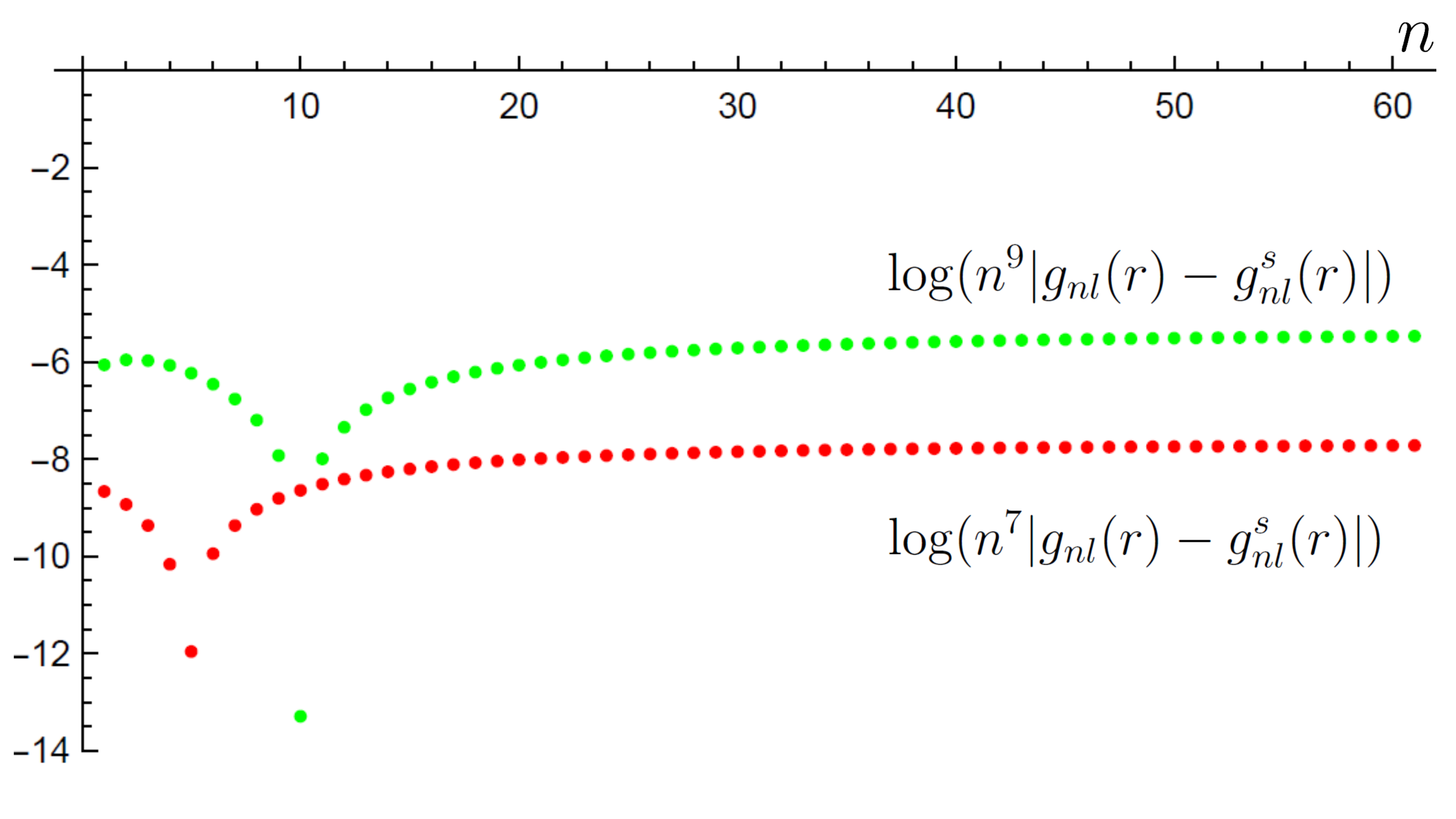}
\caption{Log plots showing convergence over $n$ in the mode sums expression. The red line represents $\log(n^7 |g_{nl}(r)-g_{nl}^{\subS}(r)|)$ where we have not included the $O(\epsilon^2)$ terms in $g_{nl}^{\subS}(r)$ (setting $m=1$). The plot shows that the difference $g_{nl}(r)-g_{nl}^{\subS}(r)$ scales like $n^{-7}$ for large $n$. The green line represents $\log(n^9|g_{nl}(r)-g_{nl}^{\subS}(r)|)$ where we have included $O(\epsilon^2)$ terms (setting $m=2$) . The plot shows that the difference scales like $n^{-9}$ for large $n$. }
\label{fig:d5nconvergence}
\end{figure}

First, in order to make the mode-by-mode subtraction more transparent, it is useful to simplify the notation by defining
\begin{align}
	g_{nl}^{\subS}(r)&=\frac{\Omega_{d-2}}{2\kappa}\frac{\Gamma(\frac{d}{2}-1)}{(2\pi)^{\frac{d}{2}-1}}\Big\{\sum_{i=0}^{\frac{d}{2}}\sum_{j=0}^{i}\mathcal{D}_{ij}^{(+)}(r)\stackrel{[d]}{\Psi}\!\!{}_{nl}^{(+)}(i,j|r)\nonumber\\
	&+\sum_{i=1}^{\frac{d}{2}}\sum_{j=1}^{\min\{i,\frac{d}{2}-2\}}\mathcal{D}_{i,j}^{(-)}(r)\stackrel{[d]}{\Psi}\!\!{}_{nl}^{(-)}(i,j|r)\nonumber\\
	&+\mathcal{T}_{10}^{(\textrm{r})}(r)\,\stackrel{[d]}{\Psi}\!\!{}_{nl}^{(-)}(\tfrac{d}{2},\tfrac{d}{2}-2|r)\nonumber\\
	&+\sum_{i=0}^{2}\sum_{j=0}^{i}\mathcal{T}_{ij}^{\textrm{(l)}}(r)\,\stackrel{[d]}{\chi}\!\!{}_{nl}(i,j|r) \Big\},
\end{align}
whence the mode-sum of the Hadamard parametrix can be more succinctly expressed as
\begin{align}
	\label{eq:modeHadamard}
	G_{\subS}(x,x')=\frac{\kappa}{2\pi}\sum_{l=0}^{\infty}\frac{(2l+2\mu)}{\Omega_{d-2}}C_{l}^{\mu}(\cos\gamma)\Bigg\{g_{0l}^{\subS}(r)\nonumber\\
	+2\sum_{n=1}^{\infty}\cos\kappa\Delta\tau\,g_{nl}^{\subS}(r)\Bigg\}+\frac{\Gamma(\frac{d}{2}-1)}{2(2\pi)^{d/2}}\mathcal{D}_{\frac{d}{2}-1,\frac{d}{2}-1}^{(-)}(r).
\end{align}
In arriving at this form we have also made use of the invariance of $g_{nl}^{\subS}(r)$ under the $n\to-n$, a symmetry that is immediately obvious from the explicit expressions for the regularization parameters. The radial part of the Euclidean Green function also possesses this discrete symmetry and can be written in an analogous way. Hence, substituting (\ref{eq:modeHadamard}) and (\ref{eq:modeGreen}) into the definiton (\ref{eq:phi2Def}), and using the fact that the Gegenbauer polynomials evaluated at coincidence are
\begin{align}
	C_{l}^{\mu}(1)=\binom{2\mu+l-1}{l},
\end{align}
yields the following expression for the regularized vacuum polarization
\begin{align}
\langle & \phi^{2} \rangle_{\textrm{reg}}=\frac{\kappa}{2\pi}\sum_{l=0}^{\infty}\frac{(2l+2\mu)}{\Omega_{d-2}}\binom{2\mu+l-1}{l}\Big\{g_{0l}(r)-g_{0l}^{\subS}(r)\nonumber\\
&	+2\sum_{n=1}^{\infty}(g_{nl}(r)-g_{nl}^{\subS}(r))\Big\}+\frac{\Gamma(\frac{d}{2}-1)}{2(2\pi)^{d/2}}\mathcal{D}_{\frac{d}{2}-1,\frac{d}{2}-1}^{(-)}(r).
\end{align}
This double-sum is now rapidly convergent and amenable to numerical evaluation. More specifically, when terms up to and including $\mathcal{O}(\epsilon^{2})$ are included in the decomposition of the Hadamard parametrix, as is the case in this paper, the convergence of the inner sum over $n$ can be shown numerically to be $\mathcal{O}(n^{-d-3})$ for each value of $d$ under consideration (see Fig. \ref{fig:d5nconvergence} for plots of convergence for $d=6$). Of course, in principle, any order of convergence can be achieved by decomposing higher and higher order terms in the Hadamard parametrix.

\begin{figure}
	\includegraphics[width=9.2cm]{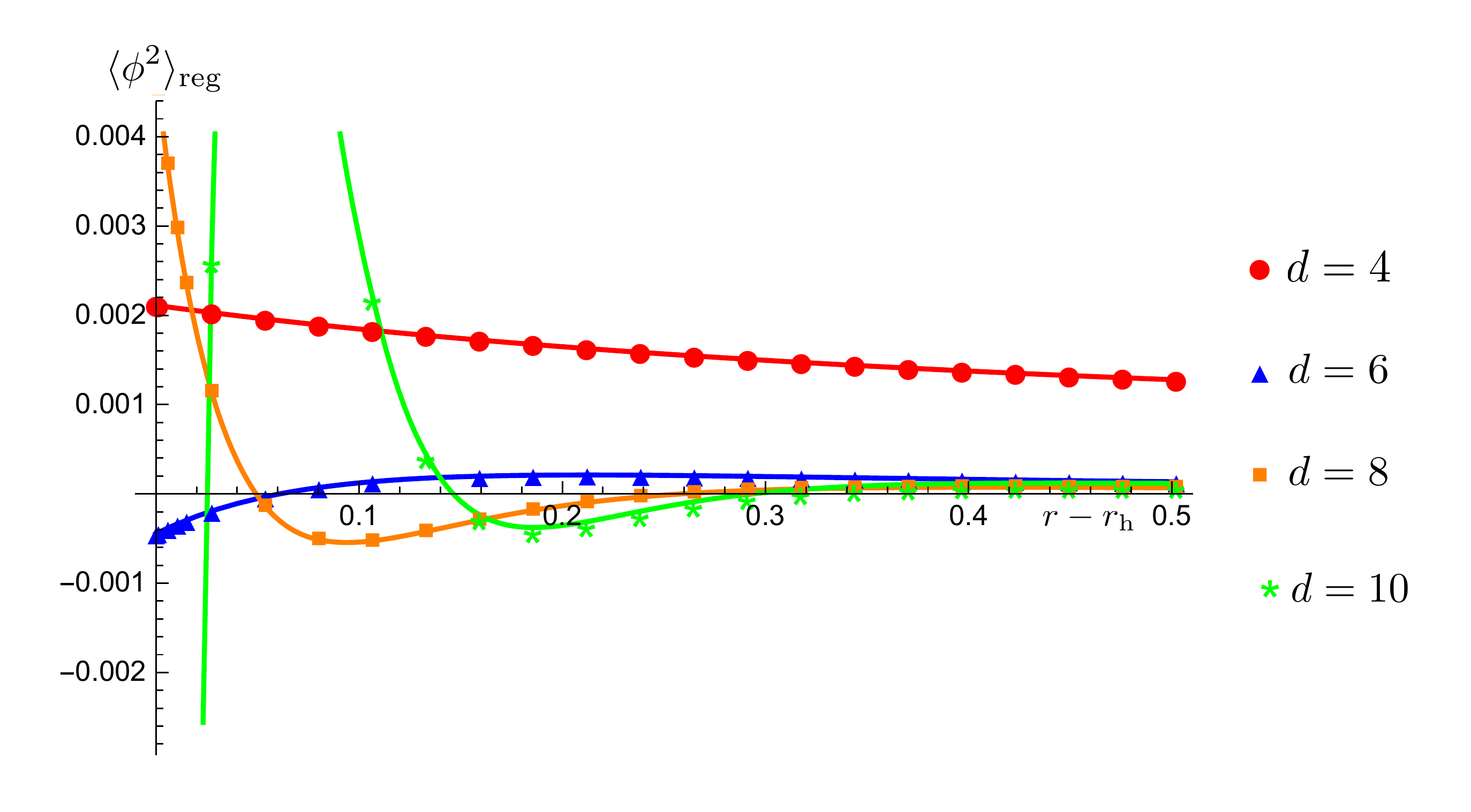}
	\caption{Plot of the regularized vacuum polarization for a massless scalar field in the Hartle-Hawking state in the exterior region of a Schwarzschild-Tangherlini black hole as a function of $r$ for spacetime dimensions $d=4,6,8,10$. We have set $r_{\textrm{h}}=\ell=1$. The plot markers represent the grid points at which $\langle \phi^2 \rangle_{\textrm{reg}}$ was numerically calculated.}
	\label{fig:phid46810}
\end{figure}

\begin{figure*}
	\includegraphics[width=16cm]{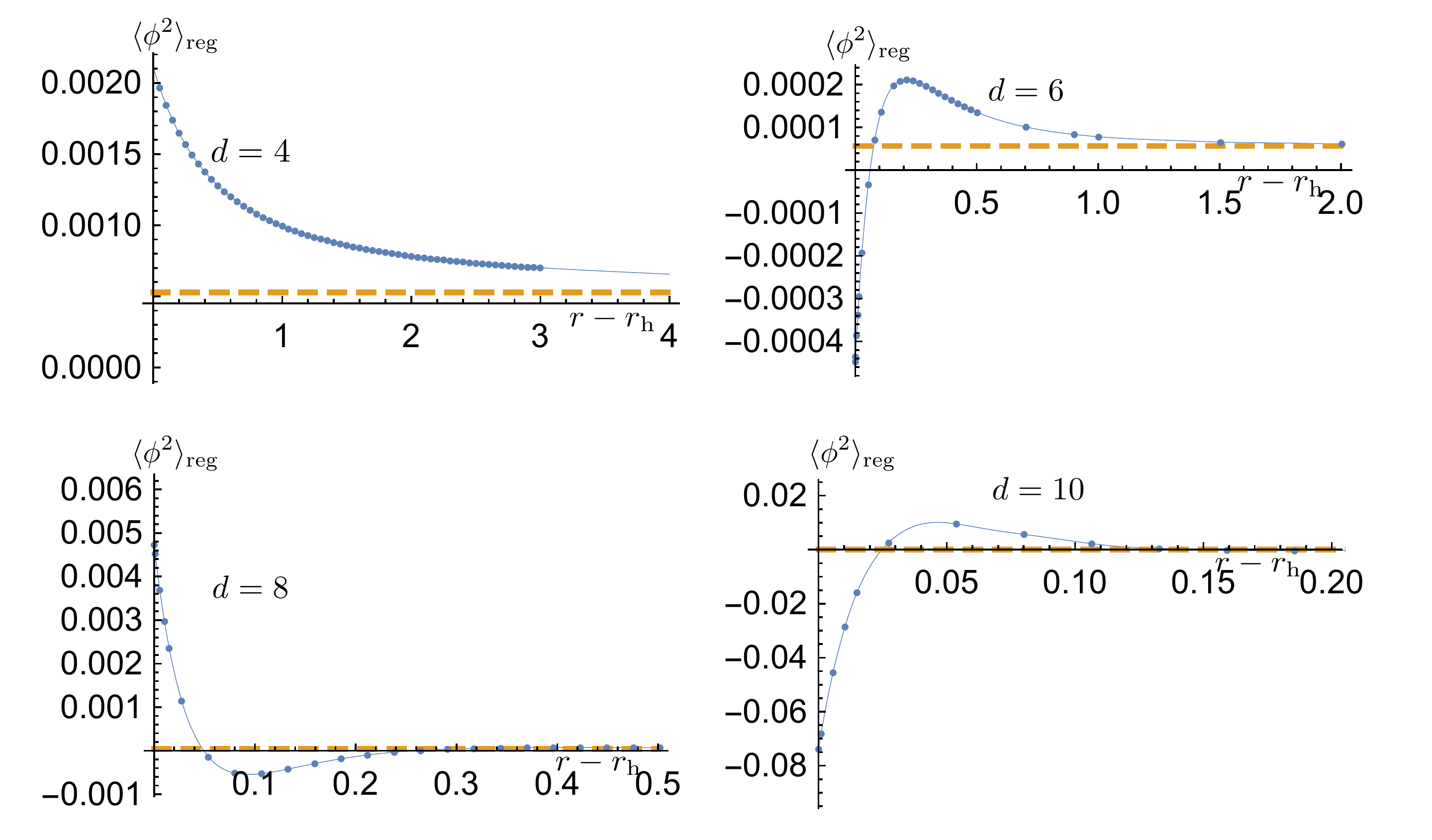}
	\caption{Separate plots of the regularized vacuum polarization for a massless scalar field in the Hartle-Hawking state in the exterior region of a Schwarzschild-Tangherlini black hole as a function of $r$ in various even dimensions. We have set $r_{\textrm{h}}=\ell=1$. For $d=6,8$ and $d= 10$ the plot markers represent the grid points at which we numerically evaluate $\langle \phi^2 \rangle_{\textrm{reg}}$, while for $d=4$ they represent the tabulated results of \cite{CandelasHowardPhi2}. Comparison of the tabulated data with the graph for $d=4$ shows excellent agreement. The dashed lines are the asymptotic values given by Eq.(\ref{eq:flatspace}).}
	\label{fig:ind}
\end{figure*}

Below, we present plots of $\langle \phi^2 \rangle_{\textrm{reg}}$ in the exterior region of a Schwarzschild-Tangherlini black hole space-time for $d=4,6,8,10$. Recall that we are working in units where the black hole event horizon has been set to unity. In Fig. \ref{fig:phid46810} we present the results for all dimensions. This is followed by a series of individual plots for each dimension, in Fig. \ref{fig:ind}. From the plots, we might conjecture that for
d = 6, 10,..., the vacuum polarization is rapidly increasing from a negative value at the horizon out to some turning point, before
decreasing and eventually approaching its value at infinity. For the alternate even dimensions 8, 12,.., the vacuum polarization decreases rapidly from a positive value at the horizon to some turning point, before slowly increasing and eventually asymptoting to its value at infinity. A similar pattern of alternating dimensions of the same parity having similar shape graphs was seen in the odd-dimensional case also \cite{TaylorBreenOdd}.

Another consistency check of our numerical implementation is to compare our graphs with the known values for the vacuum polarization at the horizon and infinity. In the latter case, the fact that Schwarzschild-Tangherlini spacetime is asymptotically flat implies that, as $r\to\infty$, $\langle \phi^2 \rangle_{\textrm{reg}}$ ought to approach the value of the regularized vacuum
polarization for a thermal scalar field in flat spacetime at the Hawking temperature $T=\kappa/2\pi$, which is given by:
\be
\label{eq:flatspace}
\langle \phi^2 \rangle^{\mathcal{M}}_{\textrm{reg}}=\frac{\kappa^{d-2}\Gamma\left(\frac{d}{2}-1\right)}{2^{d-1}\pi^{3d/2-2}} \zeta(d-2),
\ee
where $\zeta(s)$ is the Riemann zeta function. This agreement as $r\to\infty$ can be seen in Fig. \ref{fig:ind}. Moreover, the rate at which $\langle \phi^2 \rangle_{\textrm{reg}}$ approaches its flat spacetime value increases with the number of spacetime dimensions. 

Turning now to the comparison with (quasi-)analytic expressions for the vacuum polarization on the horizon. In many cases, the Green function near the horizon, and hence the vacuum polarization, can be computed in closed-form since only the $n=0$ mode contributes. Even in the cases where explicit closed-form representations are not available, one can derive simple integral representations for the Green function by extending the work of \cite{Fro:89}. Moreover, analogous integral representations for the vacuum polarization can be obtained from these which are straightforward to evaluate numerically. Computing vacuum polarization on the horizon in this independent way and checking that these values smoothly match with our first off-horizon values calculated using the methods described in this paper offers another non-trivial check of the validity of our results. In our plots of  $\langle \phi^2 \rangle_{\textrm{reg}}$, the first grid point is taken to be the value of $\langle \phi^2 \rangle_{\textrm{reg}}$ exactly on the black hole horizon. The relevant horizon values are given in Table \ref{tab:hor} below (with $r_h=1$).
\begin{table}[!ht]
	\begin{tabular}{|c|c|}
		\hline
		$d$&$\langle \phi^2\rangle_{\textrm{reg}}$ at $r=r_{\textrm{h}}=1$\\
		\hline
				&\\
		4&$\displaystyle{\frac{1}{48 \pi ^2}}$\\ 
		&\\
		6&$\displaystyle{\frac{2 \gamma +\ln (3)}{48 \pi ^3}-\frac{21}{320 \pi ^3}}$\\ 
			&\\
		8&  0.00539968702\\
			&\\
		10&-0.08070202480\\
		\hline
	\end{tabular}
	\caption{Values of the renormalized vacuum polarization on the event horizon of a Schwarzschild-Tangherlini black hole in various even dimensions from $d=4,..,10$. The event horizon is located at $r=r_{\textrm{h}}=1$.}
	\label{tab:hor}
\end{table}
While the result for $d=4$ is well-known \cite{Candelas:1980zt} and straightforward to derive explicitly in closed form, to the best of the authors' knowledge this is the first instance where a closed form result for $d=6$ is given. The results for $d=8$ and $d=10$ were calculated numerically using integral expressions that generalize the one given in \cite{Fro:89}. The important thing to note is that in each of the plots in Fig.(\ref{fig:ind}), the value of $\langle \phi^2 \rangle_{\textrm{reg}}$ at the first off-horizon grid point (which is calculated using the new method described in this paper) matches up smoothly with the horizon value (which is calculated using an independent method). More than simply a check of the validity of our method, this smooth matching further exhibits little or no loss of accuracy/efficiency of the mode-sums very near the horizon, that is, the method developed in this series of papers results in mode-sums whose convergence properties are excellent across the entire exterior region. This desirable property is not shared by the usual methods based on WKB techniques, where the convergence close to the horizon breaks down. We feel that this uniformity is a major advantage of our method.

\begin{figure*}
	\includegraphics[width=15cm]{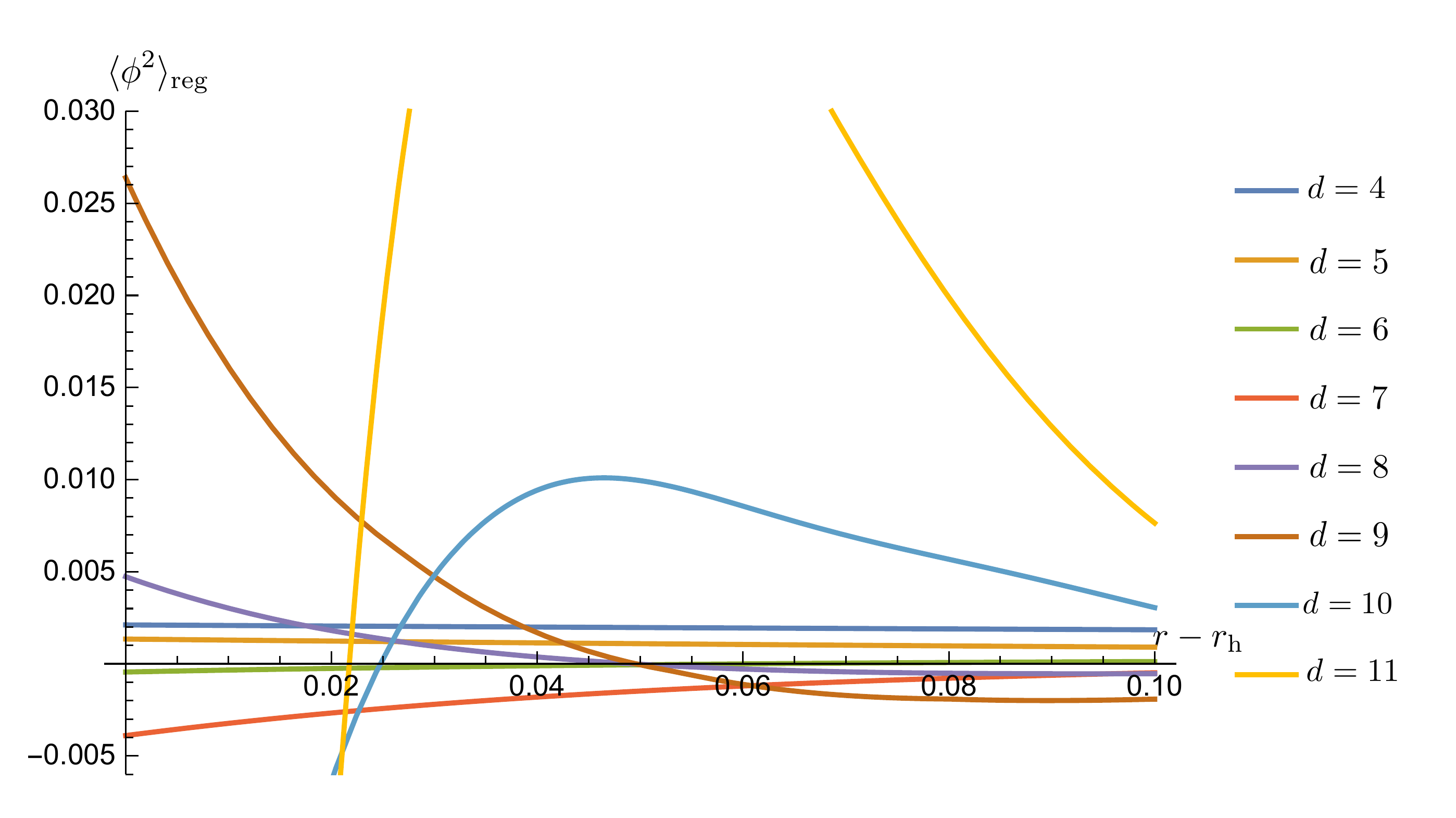}
	\caption{Plot of the regularized vacuum polarization for a massless scalar field in the Hartle-Hawking state in the exterior region of a Schwarzschild-Tangherlini black hole for spacetime dimensions $d=4,...,11$. We have set $r_{\textrm{h}}=\ell=1$.}
	\label{fig:phiall}
	\end{figure*}

Finally in Fig (\ref{fig:phiall}), we draw together the results from this series of papers to present the vacuum polarization in the exterior region of a Schwarzschild-Tangherlini black hole spacetime for all spacetime dimensions $d=4,...,11$. We see that for $d=4$ and $d=5$  $\langle \phi^2 \rangle_{\textrm{reg}}$  remains positive in the entire exterior region, for $d=6$ and $d=7$ $\langle \phi^2 \rangle_{\textrm{reg}}$ is negative at the event horizon, it then increases to a positive maximum value before limiting to its asymptotic value. For $d=8$ and $d=9$, the vacuum polarization is positive on the event horizon, it then decreases to a minimum negative value before increasing again and then approaches its  asymptotic value. Lastly, for $d=10$ and $d=11$, $\langle \phi^2 \rangle_{\textrm{reg}}$ is negative on the horizon, it then increases to a maximum positive value before decreasing and becoming slightly negative before limiting to its flat space value. This grouping between neighboring dimensions appears to be a universal feature, though it is unclear if it is a physically interesting one. If this type of pairing persisted in the calculation of the stress-energy tensor, perhaps it would be worth further investigation.

\section{Conclusions}
We have extended the method presented in Ref.~\cite{TaylorBreenOdd} for computing vacuum polarization in odd dimensions for a quantum scalar field in the Hartle-Hawking state in static, spherically-symmetric spacetimes to the even-dimensional case. These methods offer extremely powerful tools for computing regularized vacuum polarization for fields propagating in static, spherically symmetric spacetimes of any dimension. Computing regularized vacuum polarization, even in four spacetime dimensions, is historically a notoriously difficult and technical task. The first successful regularization calculation of this type was presented by Candelas and Howard for a scalar field in a Schwarzschild black hole spacetime. Their method is quite ingenious, relying on the application of WKB techniques and converting one of the infinite series to a contour integral using tools from complex analysis. However, after much endeavor and artfulness, one is still left with expressions which are inefficient to compute numerically. Moreover, the method fails completely near the horizon. Despite these drawbacks, this approach has more-or-less remained the standard one for several decades. This is remedied here by presenting a systematic out-of-the-box solution that is more direct, conceptually clearer and much more efficient than the Candelas Howard approach. Moreover, the approach presented here is mostly agnostic to number of dimensions, to the mass of the field or to whether or not the spacetime is vacuum. Our approach results in a mode-by-mode subtraction for the vacuum polarization that is very rapidly converging, requiring only a few tens of $l$ and $n$ modes to obtain approximately 8-10 decimal places of accuracy on the entire exterior geometry. Finally, we conclude that, in principle, these methods can be extended to the regularized stress-energy tensor, to other types of fields and to other vacuum states.

 \begin{table*}
 	\begin{center}
 	\begin{tabular}{|c|ccccc|} \hline
 		& \multicolumn{5}{c|}{$\mathcal{D}^{(+)}_{ij}(r)$ coefficients for 6D Schwarzschild-Tangherlini}\\ \hline \hline
 		& $j=0$ & $j=1$ & $j=2$ & $j=3$ & \\ \hline &&&&&\\
 $i=0$& $4$ & \text{} & \text{} & \text{} & \text{} \\ &&&&&\\
 $i=1$& $\dfrac{10 \left(r^3-1\right)}{3
    r^8}$ & $\begin{array}{c}-\dfrac{1}{6
    r^{11}}\Big(9 r^{11}-9 r^8\\+16 r^6-41 r^3+25\Big)\end{array}$ &
    \text{} & \text{} &
    \text{} \\ & &&&&\\
 $i=2$& $\begin{array}{c}\dfrac{1}{12 r^{16}}\Big(9 r^{11}-18 r^9-9 r^8\\+112
    r^6-179 r^3+85\Big)\end{array}$ & $\begin{array}{c}\dfrac{1}{720 r^{19}}\Big(-324 r^{19}+324
    r^{16}\\-1620 r^{14}+480 r^{12}+3645 r^{11}\\-5216 r^9-2025
    r^8+15036 r^6\\-16425 r^3+6125\Big)\end{array}$ &
    $\begin{array}{c}\dfrac{(r-1)^{4}}{192 r^{22}}
    \Big(9 r^9+18 r^8\\+27 r^7+27 r^6+27 r^5\\+43 r^4+59 r^3+75
    r^2\\+50 r+25\Big)^{2}\end{array}$ &
    $$ & $$
    \\ &&&&&\\
$i=3$ & $\begin{array}{c}\dfrac{1}{8640 r^{24}}\Big(1944
    r^{19}-6480 r^{17}\\-1944 r^{16}+4680 r^{15}\\+36315
    r^{14}-69540 r^{12}\\-55620 r^{11}+312416 r^9\\+25785
    r^8-587571 r^6\\+493140 r^3-153125\Big)\end{array}$
	 & $\begin{array}{c}-\dfrac{1}{161280 r^{27}}\Big(26244
   r^{27}-26244 r^{24}\\+165564 r^{22}-151200 r^{20}\\-366849
   r^{19}+26880 r^{18}\\+1312416 r^{17}+201285 r^{16}\\-649600
   r^{15}-3361176 r^{14}\\+4320736 r^{12}+3400110
   r^{11}\\-12376896 r^9-1200150 r^8\\+17417980 r^6-11902225
   r^3\\+3163125\Big)\end{array}$ & $\begin{array}{c}\dfrac{1}{3840 r^{30}}\Big(972 r^{30}-1944
   r^{27}\\+5238 r^{25}+972 r^{24}\\-1440 r^{23}-17901
   r^{22}\\+20928 r^{20}+20088 r^{19}\\-2560 r^{18}-80436
   r^{17}\\-7425 r^{16}+30112 r^{15}\\+131598 r^{14}-129344
   r^{12}\\-98400 r^{11}+273342 r^9\\+27750 r^8-306925 r^6\\+176000
   r^3-40625\Big)\end{array}$ & $\begin{array}{c}-\dfrac{(r-1)^6
   (r^2+r+1)^3} {6912 r^{33}}\Big(9 r^7\\+9 r^6+9 r^5+9 r^4+9
   r^3\\+25 r^2+25 r+25\Big)^3\end{array}$ &\\ \hline
    \end{tabular}
    \caption{We list the Hadamard direct coefficients $\mathcal{D}_{ij}^{(+)}(r)$ for the $d=6$ Schwarzschild-Tangherlini spacetime. In the table, we list all coefficients needed in the expansion of the Hadamard parametrix assuming we keep all terms up to and including $\mathcal{O}(\epsilon^{2})$. The horizon radius has been set to unity.}
    \label{tab:Dcoeffplus}
    \end{center}
 \end{table*}

 \begin{table*}[h]
 	\begin{center}
 	\begin{tabular}{|c|cccc|} \hline
 		& \multicolumn{4}{c|}{$\mathcal{D}^{(-)}_{ij}(r)$ coefficients for 6D Schwarzschild-Tangherlini}\\ \hline \hline
 		& $j=1$ & $j=2$ & $j=3$ &  \\ \hline &&&&\\
 $i=1$& $-\dfrac{2}{3 r^5}$ & \text{} & \text{} & \text{}  \\ &&&&\\
 $i=2$& $\dfrac{4 r^6-11
   r^3+7}{4 r^{13}}$ & $\dfrac{59-30 r^3}{180
   r^{10}}$ &
    \text{} &  \\ &&&&\\
 $i=3$& $\begin{array}{c}\dfrac{1}{576
   r^{21}}\Big(108 r^{14}-360 r^{12}-297
   r^{11}+3828 r^9\\+189 r^8-10520 r^6+11077 r^3-4025\Big)\end{array}$ & $\begin{array}{c}\dfrac{\left(r^3-1\right) \left(588 r^6-3065
   r^3+3325\right)}{2016 r^{18}}\end{array}$ &
    $\begin{array}{c}-\dfrac{42 r^6-325
   r^3+321}{1008 r^{15}}\end{array}$ &
     \\ \hline
    \end{tabular}
    \caption{We list the Hadamard direct coefficients $\mathcal{D}_{ij}^{(-)}(r)$ for the $d=6$ Schwarzschild-Tangherlini spacetime. In the table, we list all coefficients needed in the expansion of the Hadamard parametrix assuming we keep all terms up to and including $\mathcal{O}(\epsilon^{2})$. The horizon radius has been set to unity.}
    \label{tab:Dcoeffminus}
    \end{center}
 \end{table*}

 \begin{table*}[h]
 	\begin{center}
 	\begin{tabular}{|c|ccc|} \hline
 		& \multicolumn{3}{c|}{$\mathcal{T}^{\textrm{(l)}}_{ij}(r)$ coefficients for }\\ 
		& \multicolumn{3}{c|}{6D Schwarzschild-Tangherlini}\\ \hline \hline 
 		& $j=0$ & $j=1$ &  \\ \hline &&&\\
 $i=0$& $-\dfrac{1}{3 r^{10}}$ & \text{} &   \\ &&&\\
 $i=1$& $\dfrac{5
   \left(3 r^3-5\right)}{36 r^{15}}$ & $-\dfrac{5
   \left(r^3-1\right)}{12 r^{15}}$ &
    
    \\ &&&\\ \hline
    \end{tabular}
	\quad
    	\begin{tabular}{|c|ccc|} \hline
    		& \multicolumn{3}{c|}{$\mathcal{T}^{\textrm{(r)}}_{ij}(r)$ and $\mathcal{T}^{\textrm{(p)}}_{ij}(r)$ coefficients for }\\ 
   		& \multicolumn{3}{c|}{6D Schwarzschild-Tangherlini}\\ \hline \hline 
    		& $j=0$ & $j=1$ &  \\ \hline &&&\\
    $\mathcal{T}_{1j}^{(\textrm{r})}$& $-\dfrac{9 r^{11}-9 r^8+16 r^6-41 r^3+25}{144 r^{21}}$ & \text{} &   \\ &&&\\
    $\mathcal{T}_{1j}^{(\textrm{p})}$& $-\dfrac{1}{36 r^{15}}$ & $\dfrac{5 \left(r^3-1\right)}{36 r^{18}}$ &
    
       \\ &&&\\ \hline
       \end{tabular}
    \caption{We list the Hadamard tail coefficients $\mathcal{T}_{ij}^{(\textrm{l})}(r)$, $\mathcal{T}_{ij}^{(\textrm{r})}(r)$ and $\mathcal{T}_{ij}^{(\textrm{p})}(r)$ for a massless scalar field in  $d=6$ Schwarzschild-Tangherlini spacetime. These are all the terms needed assuming we keep terms in the Hadamard expansion up to and including $\mathcal{O}(\epsilon^{2})$ terms. In $d=4$, for a massless scalar field in Schwarzschild, we have that $V\sim \mathcal{O}(\epsilon^{4})$, and hence all of the $\mathcal{T}_{ij}$ terms would be zero at this order. As we can see, this is not true in the higher dimensional Schwarzschild-Tangherlini spacetimes.}
    \label{tab:Tcoeff}
    \end{center}
 \end{table*}
 \vfill
\bibliographystyle{apsrev4-1}
\bibliography{my_bib}

\begin{thebibliography}{18}%
\makeatletter
\providecommand \@ifxundefined [1]{%
 \@ifx{#1\undefined}
}%
\providecommand \@ifnum [1]{%
 \ifnum #1\expandafter \@firstoftwo
 \else \expandafter \@secondoftwo
 \fi
}%
\providecommand \@ifx [1]{%
 \ifx #1\expandafter \@firstoftwo
 \else \expandafter \@secondoftwo
 \fi
}%
\providecommand \natexlab [1]{#1}%
\providecommand \enquote  [1]{``#1''}%
\providecommand \bibnamefont  [1]{#1}%
\providecommand \bibfnamefont [1]{#1}%
\providecommand \citenamefont [1]{#1}%
\providecommand \href@noop [0]{\@secondoftwo}%
\providecommand \href [0]{\begingroup \@sanitize@url \@href}%
\providecommand \@href[1]{\@@startlink{#1}\@@href}%
\providecommand \@@href[1]{\endgroup#1\@@endlink}%
\providecommand \@sanitize@url [0]{\catcode `\\12\catcode `\$12\catcode
  `\&12\catcode `\#12\catcode `\^12\catcode `\_12\catcode `\%12\relax}%
\providecommand \@@startlink[1]{}%
\providecommand \@@endlink[0]{}%
\providecommand \url  [0]{\begingroup\@sanitize@url \@url }%
\providecommand \@url [1]{\endgroup\@href {#1}{\urlprefix }}%
\providecommand \urlprefix  [0]{URL }%
\providecommand \Eprint [0]{\href }%
\providecommand \doibase [0]{http://dx.doi.org/}%
\providecommand \selectlanguage [0]{\@gobble}%
\providecommand \bibinfo  [0]{\@secondoftwo}%
\providecommand \bibfield  [0]{\@secondoftwo}%
\providecommand \translation [1]{[#1]}%
\providecommand \BibitemOpen [0]{}%
\providecommand \bibitemStop [0]{}%
\providecommand \bibitemNoStop [0]{.\EOS\space}%
\providecommand \EOS [0]{\spacefactor3000\relax}%
\providecommand \BibitemShut  [1]{\csname bibitem#1\endcsname}%
\let\auto@bib@innerbib\@empty
%</preamble>
\bibitem [{\citenamefont {DeWitt}(1975)}]{dewitt1975quantum}%
  \BibitemOpen
  \bibfield  {author} {\bibinfo {author} {\bibfnamefont {B.~S.}\ \bibnamefont
  {DeWitt}},\ }\href@noop {} {\bibfield  {journal} {\bibinfo  {journal}
  {Physics Reports}\ }\textbf {\bibinfo {volume} {19}},\ \bibinfo {pages} {295}
  (\bibinfo {year} {1975})}\BibitemShut {NoStop}%
\bibitem [{\citenamefont {Christensen}(1976)}]{ChristensenPointSplit}%
  \BibitemOpen
  \bibfield  {author} {\bibinfo {author} {\bibfnamefont {S.~M.}\ \bibnamefont
  {Christensen}},\ }\href {\doibase 10.1103/PhysRevD.14.2490} {\bibfield
  {journal} {\bibinfo  {journal} {Phys. Rev. D}\ }\textbf {\bibinfo {volume}
  {14}},\ \bibinfo {pages} {2490} (\bibinfo {year} {1976})}\BibitemShut
  {NoStop}%
\bibitem [{\citenamefont {Candelas}\ and\ \citenamefont
  {Howard}(1984)}]{CandelasHowardPhi2}%
  \BibitemOpen
  \bibfield  {author} {\bibinfo {author} {\bibfnamefont {P.}~\bibnamefont
  {Candelas}}\ and\ \bibinfo {author} {\bibfnamefont {K.~W.}\ \bibnamefont
  {Howard}},\ }\href {\doibase 10.1103/PhysRevD.29.1618} {\bibfield  {journal}
  {\bibinfo  {journal} {Phys. Rev. D}\ }\textbf {\bibinfo {volume} {29}},\
  \bibinfo {pages} {1618} (\bibinfo {year} {1984})}\BibitemShut {NoStop}%
\bibitem [{\citenamefont {Taylor}\ and\ \citenamefont
  {Breen}(2016)}]{TaylorBreenOdd}%
  \BibitemOpen
  \bibfield  {author} {\bibinfo {author} {\bibfnamefont {P.}~\bibnamefont
  {Taylor}}\ and\ \bibinfo {author} {\bibfnamefont {C.}~\bibnamefont {Breen}},\
  }\href {\doibase 10.1103/PhysRevD.94.125024} {\bibfield  {journal} {\bibinfo
  {journal} {Phys. Rev. D}\ }\textbf {\bibinfo {volume} {94}},\ \bibinfo
  {pages} {125024} (\bibinfo {year} {2016})}\BibitemShut {NoStop}%
\bibitem [{\citenamefont {Morgan}\ \emph {et~al.}(2007)\citenamefont {Morgan},
  \citenamefont {Thom}, \citenamefont {Winstanley},\ and\ \citenamefont
  {Young}}]{Morgan:2007hp}%
  \BibitemOpen
  \bibfield  {author} {\bibinfo {author} {\bibfnamefont {D.}~\bibnamefont
  {Morgan}}, \bibinfo {author} {\bibfnamefont {S.}~\bibnamefont {Thom}},
  \bibinfo {author} {\bibfnamefont {E.}~\bibnamefont {Winstanley}}, \ and\
  \bibinfo {author} {\bibfnamefont {P.~M.}\ \bibnamefont {Young}},\ }\href
  {\doibase 10.1007/s10714-007-0486-3} {\bibfield  {journal} {\bibinfo
  {journal} {Gen.~Rel.~Grav.}\ }\textbf {\bibinfo {volume} {39}},\ \bibinfo
  {pages} {1719} (\bibinfo {year} {2007})},\ \Eprint
  {http://arxiv.org/abs/0705.1131} {arXiv:0705.1131 [gr-qc]} \BibitemShut
  {NoStop}%
%%CITATION = ARXIV:0705.1131;%%
\bibitem [{\citenamefont {D\'ecanini}\ and\ \citenamefont
  {Folacci}(2008)}]{DecaniniFolacciHadamardRen}%
  \BibitemOpen
  \bibfield  {author} {\bibinfo {author} {\bibfnamefont {Y.}~\bibnamefont
  {D\'ecanini}}\ and\ \bibinfo {author} {\bibfnamefont {A.}~\bibnamefont
  {Folacci}},\ }\href {\doibase 10.1103/PhysRevD.78.044025} {\bibfield
  {journal} {\bibinfo  {journal} {Phys. Rev. D}\ }\textbf {\bibinfo {volume}
  {78}},\ \bibinfo {pages} {044025} (\bibinfo {year} {2008})}\BibitemShut
  {NoStop}%
\bibitem [{\citenamefont {Thompson}\ and\ \citenamefont
  {Lemos}(2009)}]{Thompson:2008bk}%
  \BibitemOpen
  \bibfield  {author} {\bibinfo {author} {\bibfnamefont {R.~T.}\ \bibnamefont
  {Thompson}}\ and\ \bibinfo {author} {\bibfnamefont {J.~P.~S.}\ \bibnamefont
  {Lemos}},\ }\href {\doibase 10.1103/PhysRevD.80.064017} {\bibfield  {journal}
  {\bibinfo  {journal} {Phys.~Rev.}\ }\textbf {\bibinfo {volume} {D 80}},\
  \bibinfo {pages} {064017} (\bibinfo {year} {2009})},\ \Eprint
  {http://arxiv.org/abs/0811.3962} {arXiv:0811.3962 [gr-qc]} \BibitemShut
  {NoStop}%
%%CITATION = ARXIV:0811.3962;%%
\bibitem [{\citenamefont {Matyjasek}\ and\ \citenamefont
  {Sadurski}(2015{\natexlab{a}})}]{MatyjasekHD1}%
  \BibitemOpen
  \bibfield  {author} {\bibinfo {author} {\bibfnamefont {J.}~\bibnamefont
  {Matyjasek}}\ and\ \bibinfo {author} {\bibfnamefont {P.}~\bibnamefont
  {Sadurski}},\ }\href {\doibase 10.1103/PhysRevD.92.044023} {\bibfield
  {journal} {\bibinfo  {journal} {Phys. Rev. D}\ }\textbf {\bibinfo {volume}
  {92}},\ \bibinfo {pages} {044023} (\bibinfo {year}
  {2015}{\natexlab{a}})}\BibitemShut {NoStop}%
\bibitem [{\citenamefont {Matyjasek}\ and\ \citenamefont
  {Sadurski}(2015{\natexlab{b}})}]{MatyjasekHD2}%
  \BibitemOpen
  \bibfield  {author} {\bibinfo {author} {\bibfnamefont {J.}~\bibnamefont
  {Matyjasek}}\ and\ \bibinfo {author} {\bibfnamefont {P.}~\bibnamefont
  {Sadurski}},\ }\href {\doibase 10.1103/PhysRevD.91.044027} {\bibfield
  {journal} {\bibinfo  {journal} {Phys. Rev. D}\ }\textbf {\bibinfo {volume}
  {91}},\ \bibinfo {pages} {044027} (\bibinfo {year}
  {2015}{\natexlab{b}})}\BibitemShut {NoStop}%
\bibitem [{\citenamefont {Wald}(1994)}]{WaldQFT}%
  \BibitemOpen
  \bibfield  {author} {\bibinfo {author} {\bibfnamefont {R.~M.}\ \bibnamefont
  {Wald}},\ }\href@noop {} {\emph {\bibinfo {title} {Quantum Field Theory in
  Curved Spacetime and Black Hole Thermodynamics}}}\ (\bibinfo  {publisher}
  {University of Chicago Press},\ \bibinfo {year} {1994})\BibitemShut {NoStop}%
\bibitem [{MyW()}]{MyWebpage}%
  \BibitemOpen
  \href@noop {} {}\bibinfo {howpublished}
  {\url{http://www.taylorexpansion.net}}\BibitemShut {NoStop}%
\bibitem [{\citenamefont {Ottewill}\ and\ \citenamefont
  {Taylor}(2010)}]{OttewillTaylorCS}%
  \BibitemOpen
  \bibfield  {author} {\bibinfo {author} {\bibfnamefont {A.~C.}\ \bibnamefont
  {Ottewill}}\ and\ \bibinfo {author} {\bibfnamefont {P.}~\bibnamefont
  {Taylor}},\ }\href {\doibase 10.1103/PhysRevD.82.104013} {\bibfield
  {journal} {\bibinfo  {journal} {Phys. Rev. D}\ }\textbf {\bibinfo {volume}
  {82}},\ \bibinfo {pages} {104013} (\bibinfo {year} {2010})}\BibitemShut
  {NoStop}%
\bibitem [{\citenamefont {Cohl}(2013)}]{CohlGegenInt}%
  \BibitemOpen
  \bibfield  {author} {\bibinfo {author} {\bibfnamefont {H.~S.}\ \bibnamefont
  {Cohl}},\ }\href {\doibase 10.1080/10652469.2012.761613} {\bibfield
  {journal} {\bibinfo  {journal} {Integral Transforms and Special Functions}\
  }\textbf {\bibinfo {volume} {24}},\ \bibinfo {pages} {807} (\bibinfo {year}
  {2013})}\BibitemShut {NoStop}%
\bibitem [{\citenamefont {Gradshteyn}\ and\ \citenamefont
  {Ryzhik}(2000)}]{GradRiz}%
  \BibitemOpen
  \bibfield  {author} {\bibinfo {author} {\bibfnamefont {I.~S.}\ \bibnamefont
  {Gradshteyn}}\ and\ \bibinfo {author} {\bibfnamefont {I.~M.}\ \bibnamefont
  {Ryzhik}},\ }\href@noop {} {\emph {\bibinfo {title} {Table of Integrals,
  Series and Products}}}\ (\bibinfo  {publisher} {Academic Press},\ \bibinfo
  {year} {2000})\BibitemShut {NoStop}%
\bibitem [{\citenamefont {Olver}(1974)}]{Olver}%
  \BibitemOpen
  \bibfield  {author} {\bibinfo {author} {\bibfnamefont {F.~W.~J.}\
  \bibnamefont {Olver}},\ }\href@noop {} {\emph {\bibinfo {title} {Asymptotics
  and Special Functions}}}\ (\bibinfo  {publisher} {Academic University
  Press},\ \bibinfo {year} {1974})\BibitemShut {NoStop}%
\bibitem [{\citenamefont {Cohl}(2010)}]{CohlDerivQ}%
  \BibitemOpen
  \bibfield  {author} {\bibinfo {author} {\bibfnamefont {H.~S.}\ \bibnamefont
  {Cohl}},\ }\href {\doibase 10.1080/10652460903445043} {\bibfield  {journal}
  {\bibinfo  {journal} {Integral Transforms and Special Functions}\ }\textbf
  {\bibinfo {volume} {21}},\ \bibinfo {pages} {581} (\bibinfo {year} {2010})},\
  \Eprint {http://arxiv.org/abs/http://dx.doi.org/10.1080/10652460903445043}
  {http://dx.doi.org/10.1080/10652460903445043} \BibitemShut {NoStop}%
\bibitem [{\citenamefont {{Frolov}}\ \emph {et~al.}(1989)\citenamefont
  {{Frolov}}, \citenamefont {{Mazzitelli}},\ and\ \citenamefont
  {{Paz}}}]{Fro:89}%
  \BibitemOpen
  \bibfield  {author} {\bibinfo {author} {\bibfnamefont {V.~P.}\ \bibnamefont
  {{Frolov}}}, \bibinfo {author} {\bibfnamefont {F.~D.}\ \bibnamefont
  {{Mazzitelli}}}, \ and\ \bibinfo {author} {\bibfnamefont {J.~P.}\
  \bibnamefont {{Paz}}},\ }\href {\doibase 10.1103/PhysRevD.40.948} {\bibfield
  {journal} {\bibinfo  {journal} {\prd}\ }\textbf {\bibinfo {volume} {40}},\
  \bibinfo {pages} {948} (\bibinfo {year} {1989})}\BibitemShut {NoStop}%
\bibitem [{\citenamefont {Candelas}(1980)}]{Candelas:1980zt}%
  \BibitemOpen
  \bibfield  {author} {\bibinfo {author} {\bibfnamefont {P.}~\bibnamefont
  {Candelas}},\ }\href {\doibase 10.1103/PhysRevD.21.2185} {\bibfield
  {journal} {\bibinfo  {journal} {Phys. Rev. D}\ }\textbf {\bibinfo {volume}
  {21}},\ \bibinfo {pages} {2185} (\bibinfo {year} {1980})}\BibitemShut
  {NoStop}%
\end{thebibliography}%

\end{document}